\documentstyle[11pt,a4]{article}
\newlength{\bredde}
\def\slash#1{\settowidth{\bredde}{$#1$}\ifmmode\,\raisebox{.15ex}{/}
\hspace*{-\bredde} #1\else$\,\raisebox{.15ex}{/}\hspace*{-\bredde} #1$\fi}

\newcommand{\beq}{\begin{equation}}
\newcommand{\eeq}{\end{equation}}
\newcommand{\bea}{\begin{eqnarray}}
\newcommand{\eea}{\end{eqnarray}}
\newcommand{\noi}{\vspace{12pt}\noindent}
\newcommand{\lG}{\raise.3ex\hbox{$\stackrel{\leftarrow}{G}$}}
\newcommand{\lU}{\raise.3ex\hbox{$\stackrel{\leftarrow}{U}$}}
\newcommand{\lP}{\raise.3ex\hbox{$\stackrel{\leftarrow}{{\cal P}}$}}
\newcommand{\leta}{\raise.3ex\hbox{$\stackrel{\leftarrow}{\eta}$}}
\newcommand{\lOmega}{\raise.3ex\hbox{$\stackrel{\leftarrow}{\Omega}$}}
\newcommand{\ldr}{\raise.3ex\hbox{$\stackrel{\leftarrow}{\delta^r}$}}
\newcommand{\ldl}{\raise.3ex\hbox{$\stackrel{\leftarrow}{\delta^l}$}}
\newcommand{\rdr}{\raise.3ex\hbox{$\stackrel{\rightarrow}{\delta^r}$}}
\newcommand{\rdl}{\raise.3ex\hbox{$\stackrel{\rightarrow}{\delta^l}$}}

\def\beqn{\begin{eqnarray}}
\def\eeqn{\end{eqnarray}}
\def\sepand{\rule{14cm}{0pt}\and}
\def\gtwid{\raise.3ex\hbox{$>$\kern-.75em\lower1ex\hbox{$\sim$}}}
\def\ltwid{\raise.3ex\hbox{$<$\kern-.75em\lower1ex\hbox{$\sim$}}}

\begin{document}
\begin{titlepage}
\title{\Large{ Algebra of Higher Antibrackets}}


\author{{\sc Klaus Bering}\\Institute for Theoretical Physics \\
Uppsala University \\ P.O. Box 803 \\ S-751 08 Uppsala\\Sweden\\
\sepand
{\sc Poul H. Damgaard}\\
The Niels Bohr Institute\\ Blegdamsvej 17\\ DK-2100 Copenhagen\\
Denmark\\
\sepand
{\sc Jorge Alfaro}\\Fac. de Fisica\\Universidad Cat\'{o}lica de Chile\\
Casilla 306, Santiago 22, Chile}

\maketitle
\vfill
\begin{abstract}We present a simplified description of higher
antibrackets, generalizations of the conventional antibracket of
the Batalin-Vilkovisky formalism. We show that these higher
antibrackets satisfy relations that are identical to those of
higher string products in non-polynomial closed string field theory.
Generalization
to the case of $Sp(2)$-symmetry is also formulated.
\end{abstract}
\vfill
\begin{flushleft}
NBI-HE-96-16 \\
UUITP-9/96\\
hep-th/9604027
\end{flushleft}
\end{titlepage}
\newpage


\setcounter{equation}{0}
\section{Introduction}
\noi
Lagrangian BRST quantization gets its most succinct formulation in the
antibracket formalism of Batalin-Vilkovisky \cite{BV}. The basic objects
of that approach, the antibracket itself and a so-called $\Delta$-operator
(to be reviewed below), turn out to belong to a general algebraic
structure that has attracted considerable attention recently, in particular
in connection with a geometric interpretation and covariant
generalizations \cite{covariant}.

\noi
The conventional antibracket of the Batalin-Vilkovisky formalism can be
viewed as being based on a 2nd-order odd differential operator $\Delta$
satisfying $\Delta^2 = 0$. In (super) Darboux coordinates it takes the
simple form \cite{BV}
\beq
\Delta = (-1)^{\epsilon_A+1}\frac{\delta^r}{\delta\phi^A}
\frac{\delta^r}{\delta\phi^*_A} ~,\label{Delta}
\eeq
where to each field $\phi^A$ one has a matching ``antifield'' $\phi^*_A$
of Grassmann parity $\epsilon(\phi^*_A) = \epsilon(\phi^A)+1$.
The antifields are conventional antighosts of the Abelian shift symmetry
that for flat functional measures leads to the most general
Schwinger-Dyson equations \cite{AD}.

\noi
Given $\Delta$ as above, one can define an odd (statistics-changing)
antibracket $(F,G)$ from the failure of $\Delta$ to act like a derivation:
\beq
\Delta(FG) = F(\Delta G) + (-1)^{\epsilon_G}(\Delta F)G
+ (-1)^{\epsilon_G}(F,G) ~. \label{abdef}
\eeq
The antibracket so defined automatically satisfies the following relations.
First, it has an exchange symmetry of the kind
\beq
(F,G) = (-1)^{\epsilon_F\epsilon_G+\epsilon_F+\epsilon_G}(G,F)~.
\label{exchange}
\eeq
It also acts like a derivation in the sense of a generalized Leibniz rule:
\begin{eqnarray}
(F,GH) &=& (F,G)H + (-1)^{\epsilon_G(\epsilon_F+1)}G(F,H) \cr
(FG,H) &=& F(G,H) + (-1)^{\epsilon_G(\epsilon_H+1)}(F,H)G~, \label{Leibniz}
\end{eqnarray}
and it satisies a Jacobi identity,
\beq
\sum_{\mbox{\rm cycl.}}(-1)^{(\epsilon_F+1)(\epsilon_H+1)}(F,(G,H))
= 0~.\label{Jacobi}
\eeq
In addition, there is a useful relation between the $\Delta$-operator and
its associated antibracket:
\beq
\Delta (F,G) = (F,\Delta G) - (-1)^{\epsilon_G}(\Delta F,G) ~.\label{dfg}
\eeq

\noi
Recently, two of the present authors \cite{AD1} showed that
the antibracket formalism is open to a natural generalization. In a
path-integral formulation, this generalization can derived by considering
general field transformations $\phi^A \to g^A(\phi',a)$, where $a^i$
represent certain collective fields \cite{AD2}. The idea is to impose
on the Lagrangian path integral the condition that certain Ward identities
are preserved throughout the quantization procedure. If one imposes the
most general set of Ward identities possible -- the Schwinger-Dyson
equations -- through an unbroken Schwinger-Dyson BRST symmetry \cite{AD3},
one can recover the antibracket formalism of Batalin and Vilkovisky by
integrating out certain ghosts $c^A$ (the antifields $\phi^*_A$ being simply
the antighosts corresponding to $c^A$). For flat functional measures
this corresponds to
local shift transformations of the fields $\phi^A$. If the measure is not
flat, or if one wishes to impose a more restricted set of Ward identities
through the BRST symmetry, the $\Delta$-operator and the associated
antibracket will differ from those of the conventional Batalin-Vilkovisky
formalism. In ref. \cite{AD} it was shown how the Batalin-Vilkovisky
$\Delta$-operator (\ref{Delta}) can be viewed as an Abelian operator
corresponding to the Abelian shift transformation
\mbox{$\phi^A \to \phi^A - a^A$}.
The analogous non-Abelian $\Delta$-operator for general transformations
$\phi^A \to g^A(\phi'^A,a)$ was derived in ref. \cite{AD2}:
\beq
\Delta G \equiv (-1)^{\epsilon_i}\left[\frac{\delta^r}{\delta\phi^A}
\frac{\delta^r}{\delta\phi^*_i}G\right]u^A_i + \frac{1}{2}(-1)^{
\epsilon_i+1}\left[\frac{\delta^r}{\delta\phi^*_j}\frac{\delta^r}{\delta
\phi^*_i}G\right]\phi^*_kU^k_{ji} ~,\label{DeltanA}
\eeq
where the $U^k_{ij}$ are the structure coefficients for the supergroup
of transformations.\footnote{Taking for convenience that the supergroup
is semi-simple, with $(-1)^{\epsilon_i}U^i_{ij}=0$.}
They are related to the field transformations
$g^A(\phi',a)$ by the relation
\beq
\frac{\delta^r u^A_i}{\delta\phi^B}u^B_j - (-1)^{\epsilon_i\epsilon_j}
\frac{\delta^r u^A_j}{\delta\phi^B}u^B_i = -u^A_k U^k_{ij} ~,
\eeq
where
\beq
u^A_i(\phi) = \left.\frac{\delta^r g^A(\phi,a)}{\delta a^i}\right|_{a=0} ~.
\eeq

\noi
The $\Delta$-operator of eq. (\ref{DeltanA}) can be shown to be nilpotent
\cite{AD2}, and it gives rise to a new
non-Abelian antibracket by use of the relation (\ref{abdef}). Explicitly,
this antibracket takes the form \cite{AD2}
\beq
(F,G) \equiv (-1)^{\epsilon_i(\epsilon_A+1)}\frac{\delta^r F}{\delta\phi^*_i}
u^A_i\frac{\delta^l G}{\delta\phi^A} - \frac{\delta^r F}{\delta\phi^A}
u^A_i\frac{\delta^l G}{\delta\phi^*_i} + \frac{\delta^r F}{\delta\phi^*_i}
\phi^*_kU^k_{ij}\frac{\delta^l G}{\delta\phi^*_j} ~.
\eeq
In ref. \cite{AD2} this non-Abelian antibracket was derived directly in
the path integral (by integrating out the ghosts $c^A$), but it can readily
be checked that it is related to the associated $\Delta$-operator
(\ref{DeltanA}) in the manner expected from (\ref{abdef}). Because this
particular non-Abelian $\Delta$-operator is of 2nd order, the corresponding
antibracket automatically satisfies all the properties (3-6).

\noi
As shown in ref. \cite{AD1}, even this non-Abelian antibracket is open
to generalizations. One first notices that the non-Abelian $\Delta$ is
nothing but the Hamiltonian BRST operator $\Omega$ of a certain
constraint algebra in an unusual representation, that of Hamiltonian
ghost momentum. Taking the most general non-Abelian BRST operator $\Omega$
of an arbitrary non-Abelian open algebra, one can then construct the
corresponding general $\Delta$-operator by going to the ghost momentum
representation \cite{AD1}. This leads naturally to the concept of
{\em higher (non-Abelian) antibrackets}. Interestingly, much of the
appropriate mathematical machinery for such a formalism already exists
in the mathematics literature \cite{Koszul,Akman}. There is also a
surprising connection between the algebra of these higher antibrackets
and that of so-called strongly homotopy Lie algebras
(for a very readable account, written for physicists, see ref.
\cite{Stasheff}), which appear in string field theory \cite{Zwiebach}.

\noi
Interest in general Batalin-Vilkovisky algebras has recently arisen
also in the context of two-dimensional topological field theory and string
theory \cite{2d}. One should expect
the higher antibrackets to play a r\^{o}le there as well \cite{Akman}.

\noi
{}From the point of view of quantization of field theories, perhaps the
most important reason for studying the algebraic structure behind
higher antibrackets comes from the expectation that even the conventional
Batalin-Vilkovisky $\Delta$-operator will be modified by higher-order
quantum
corrections originating from operator-ordering ambiguities in the
Hamiltonian framework.\footnote{We are grateful to I.A. Batalin for
explaining this aspect to us.} This obviously makes it important to study
the Master Equation for arbitrary higher-order $\Delta$-operators, and
to understand their associated BRST structure. 

\noi
The purpose of the present paper is partly to present a simplified
construction of the higher antibrackets introduced in ref. \cite{AD1},
partly to show how they can be generalized in a natural manner to a
situation in which one has simultaneous BRST and anti-BRST symmetry.
In fact, these two symmetries can, not surprisingly, be combined into
an $Sp(2)$-symmetry. The mathematical analogue of this is
an $Sp(2)$-covariant strongly homotopy Lie algebra. While this algebra
may be of interest in its own right, it also points towards the
existence of an $Sp(2)$ BRST--anti-BRST symmetric version of closed
string field theory, as we shall show towards the end of our paper.
This will then provide a comprehensive setting for the possible
generalizations of the usual Batalin-Vilkovisky quantization formalism,
and its $Sp(2)$ extensions.

\noi
We start in section 2 with a brief review of how higher antibrackets
naturally arise if one generalizes the Batalin-Vilkovisky formalism
from shift symmetries (which generate the usual Batalin-Vilkovisky
$\Delta$-operator) to more general transformations. This is only to
set the stage for what follows, because we are in this paper interested in the
study of the higher antibrackets independently of such considerations.
We then proceed to a discussion of the Koszul construction of higher
brackets and antibrackets based on general differential operators
$\Delta$ (section 2.1). Some useful mathematical background is introduced
in section 2.2, and we
show how to reformulate this construction in a simple fashion. In
section 2.3 we discuss the precise connection to strongly homotopy
Lie algebras, and prove a useful lemma related to the algebra of
{\em two} sets of higher brackets. As an explicit realization in terms
of chosen coordinates, we describe the algebra by means of a suitable
vector field in section 2.4. The analogue of the strongly homotopy Lie
algebra structure associated with our generalized higher brackets is
discussed in section 2.6.
Section 2.5 is our first return to
physics applications: we discuss the definition of a generalized
Master Equation, first introduced in ref. \cite{AD1}. This leads us
to the subject of BRST symmetry in this higher-antibracket framework.
When formulated as the possibility of deforming a given solution of
the Master Equation by the addition of BRST-exact terms, it is of interest
to find the associated symmetry algebra. While the most
simple choice of symmetry transformations corresponds to an algebra that
is open, we show how in a simple manner one can add ``equation of
motion terms'' to the transformations in order to make the
algebra close. We also discuss finite symmetry transformations. In
section 3 we turn our attention to some intriguing parallels between
higher antibrackets and the so-called ``string products'' in closed
string field theory \cite{Kugo,Zwiebach,Sen}, when as $\Delta$-operator 
one takes the BRST charge $Q$. Section
4 is devoted to the construction of an $Sp(2)$-symmetric analogue
of the higher-antibracket BRST symmetry. Section 5 contains
our conclusions. Finally, in two appendices we propose some generalizations
which lie slightly outside the main line of the paper. In the first
(Appendix A), we show how one can introduce yet higher levels
of generalizations of the higher antibrackets discussed in the main text.
While their r\^{o}le in physics applications is totally obscure, we
nevertheless find it interesting that such a further generalization is
possible. In Appendix B we discuss generalizations of the so-called
``main identities'', valid already at the level of the normal higher
antibrackets. These new identities contain new information in cases
where, for example, $\Delta$ is no longer nilpotent, or, as discussed
in section 4, when one imposes an $Sp(2)$ symmetry as well.

\setcounter{equation}{0}
\section{Higher Antibrackets}
\noi
As explained in ref. \cite{AD1}, one can introduce obvious generalizations
of the Batalin-Vilkovisky $\Delta$-operator by considering the most
general Hamiltonian BRST operator $\Omega$ in the ghost momentum
representation. Start with a representation of first class constraints
\beq
[\lG_i,\lG_j] ~=~ i\lG_kU^k_{ij}
\eeq
of the form\cite{Batalin}
\beq
\lG_i ~\equiv~ -i\frac{\ldr}{\delta\phi^A}u^A_i ~,
\eeq
which involves a right-derivative acting to the left. Because the constraints
in this representation act to the left, one must choose a representation of
the Hamiltonian ghost (super) Heisenberg algebra
\beq
[\eta^i,{\cal P}_j] = \eta^i{\cal P}_j - (-1)^{(\epsilon_i+1)(\epsilon_j
+1)}{\cal P}_j\eta^i = i \delta^i_j
\eeq
which also involves operators acting to the left. In the ghost momentum
representation, this is
\beq
{\leta}^j = i(-1)^{\epsilon_j}\frac{\ldr}{\delta{\cal P}_j} ~.
\eeq

\noi
One of the observations in ref. \cite{AD1} is that to pass to the Lagrangian
$\Delta$-operator, one identifies the Hamiltonian ghost ${\cal P}_j$ with
the Lagrangian antighost (``antifield'') $\phi^*_j$. The most general
Hamiltonian BRST operator $\Omega$ \cite{BF}, in this representation takes
the form \cite{AD1}
\beq
\lOmega = (-1)^i\frac{\ldr}{\delta\phi^A}u^A_i\frac{\ldr}{\delta\phi^*_i}
+ \sum_{n=1}^{\infty}\phi^*_{i_{n}}\cdots
\phi^*_{i_{1}} \lU^{i_{1}\cdots i_{n}} ~,\label{leftOmega}
\eeq
where
\beq
\lU^{i_{1}\cdots i_{n}} = \frac{(-1)^{\epsilon^{i_{1}\cdots i_{n-1}}_{j_{1}
\cdots j_{n}}}}{(n+1)!}(i)^{n+1}(-1)^{\epsilon_{j_{1}} + \cdots
+ \epsilon_{j_{n+1}}}U^{i_{1}\cdots i_{n}}_{j_{1}
\cdots j_{n+1}}\frac{\ldr}{\delta\phi^*_{j_{n+1}}}\cdots\frac{\ldr}{
\delta\phi^*_{j_{1}}} ~.
\eeq
The functions $U^{i_{1}\cdots i_{n}}_{j_{1}\cdots j_{n+1}}$
are generalized structure ``constants'' of the possibly open algebra. The
infinite sum in eq. (\ref{leftOmega}) may terminate at finite order. For
example, for ordinary super Lie algebras where the structure coefficients
$U^k_{ij}$ are just constant supernumbers, the series terminates at the
first term.

\noi
The $\Delta$-operator is now defined through
\beq
\Delta F ~\equiv~ F\lOmega ~. \label{Delta-Omega}
\eeq
One immediate consequence of the fact that the quantized Hamiltonian
BRST operator satisfies $[\Omega,\Omega] = 2\Omega^2 = 0$, is that
$\Delta$ also is nilpotent. One sees that in the case of
an ordinary non-Abelian Lie algebra the general definitions (\ref{leftOmega})
and (\ref{Delta-Omega}) reproduce the $\Delta$-operator of eq.
(\ref{DeltanA}). The ordinary Batalin-Vilkovisky
formalism corresponds to Abelian shift transformations
\beq
\lG_A ~=~ -i\frac{\ldr}{\delta\phi^A}~,
\eeq
for which the general definitions (\ref{leftOmega}) and (\ref{DeltanA})
lead to the usual Batalin-Vilkovisky $\Delta$-operator (\ref{Delta}).

\noi
These preliminary remarks only serve as to motivate the study of
higher-order $\Delta$-operators, and their associated antibrackets. They show
that such higher-order $\Delta$-operators exist in the field theory
context, and can be defined by a
natural generalization of the Batalin-Vilkovisky $\Delta$-operator. But in
what follows we shall neither make explicit use of the form (\ref{leftOmega}),
nor of the precise manner in which it gives rise to new higher-order
$\Delta$-operators.

\subsection{\sc The Koszul Construction}

\noi
In this subsection, let $\Delta$ denote a Grassmann-odd differential operator
with the properties
\beq
\Delta^2 ~=~ 0 ~~,~~~~\Delta({\bf 1}) ~=~ 0 ~.\label{Delta2}
\eeq
Motivated by the previous examples, we assume that $\Delta$ differentiates
from the right. In physics, one will normally not need the case where
$\Delta({\bf 1}) \neq 0$, but exceptions exist, and these cases can be treated
with equal ease (see below). One can also relax the condition of
nilpotency without encountering difficulties.

\noi
Following Koszul \cite{Koszul}, one can define a unique antibracket
$(F,G)$, even when $\Delta$ is not of 2nd order. This is the
content of eq. (\ref{abdef}), which holds in all generality. The antibracket
so defined is a measure of the failure of $\Delta$ to act like a graded
derivation. This antibracket will automatically satisfy the exchange
relation (\ref{exchange}). The relation (\ref{dfg}) also holds in all
generality. But in general both the Leibniz rule (\ref{Leibniz}) and
the Jacobi identity (\ref{Jacobi}) will be violated.

\noi
Koszul suggests that the antibracket derived from eq. (\ref{abdef})
be used to define a ``three-bracket'', which measures the failure of the
antibracket $(F,G)$ to act like a derivation. This construction can proceed
in an iterative way to define higher and higher antibrackets. We use the
notation of ref. \cite{Koszul}, and introduce objects $\Phi^n_{\Delta}$
which are
directly related to the higher antibrackets. The lowest antibracket,
the ``one-bracket'' is essentially identified with the $\Delta$-operator
itself\footnote{An extra sign factor appears because our $\Delta$-operator
is based on right-derivatives. To facilitate a comparison with the
definitions of Koszul \cite{Koszul}, we choose to compensate explicitly
for the fact that our $\Delta$ operator is based on right derivatives. This
causes some additional sign factors in the subsequent equation.},
while the higher antibrackets can be derived from it. In detail,
\beq
\begin{array}{rcl}
\Phi^1_{\Delta}(A) &=& (-1)^{\epsilon_A}\Delta(A) \cr
\Phi^2_{\Delta}(A,B) &=& (-1)^{\epsilon_A+\epsilon_B}\Delta(AB)
- (-1)^{\epsilon_A}\Delta(A)B - (-1)^{\epsilon_A+\epsilon_B}A\Delta(B) \cr
\Phi^3_{\Delta}(A,B,C) &=& (-1)^{\epsilon_A+\epsilon_B+\epsilon_C}
\Delta(ABC) - (-1)^{\epsilon_A+\epsilon_B+\epsilon_C}A\Delta(BC) -
(-1)^{\epsilon_A+\epsilon_B}\Delta(AB)C
\cr &&+ (-1)^{\epsilon_A+\epsilon_B}A\Delta(B)C
- (-1)^{\epsilon_A+\epsilon_B+\epsilon_C+\epsilon_A\epsilon_B}B\Delta(AC)
+ (-1)^{\epsilon_B(\epsilon_A+1)
+\epsilon_A}B\Delta(A)C \cr &&+ (-1)^{\epsilon_A+\epsilon_B+\epsilon_C}AB
\Delta(C) \cr
\vdots ~~~ && ~~~\vdots \cr
\end{array}
\label{Phidef1}
\eeq
All higher antibrackets are Grassmann-odd in the sense that
\beq
\epsilon\{\Phi^n_{\Delta}(A_1,\ldots,A_n)\} = \sum_{i=1}^n \epsilon_{A_{i}}
+1 ~,
\eeq
and they satisfy a simple exchange relation:
\beq
\Phi^n_{\Delta}(A_i,\ldots,A_{i-1},A_i,\ldots,A_n) = (-1)^{\epsilon_{A_{i-1}}
\epsilon_{A_i}}\Phi^n_{\Delta}(A_i,\ldots,A_{i},A_{i-1},\ldots,A_n) ~.
\eeq
This latter relation suggests that it is more natural to view the comma in
$\Phi^n_{\Delta}$ as a graded (supercommutative) and associative product.
We use this product notation in the next sections.

\noi
The usual antibracket of the Batalin-Vilkovisky formalism,
the ``two-bracket'', is defined by
\beq
(A,B) \equiv (-1)^{\epsilon_A}\Phi^2_{\Delta}(A,B) ~.
\eeq
Note that when the usual antibracket acts like a graded derivation, the
``three-bracket'' defined through $\Phi^3_{\Delta}$ vanishes identically.

\noi
Akman \cite{Akman} has organized the above definition of higher antibrackets
in a very convenient iterative sequence:
\beq
\begin{array}{rcl}
\Phi^1_{\Delta}(A) &=& (-1)^{\epsilon_A}\Delta(A) \cr
\Phi^2_{\Delta}(A,B) &=& \Phi^1_{\Delta}(AB) - \Phi^1_{\Delta}(A) B
- (-1)^{\epsilon_A} A\Phi^1_{\Delta}(B) \cr
\Phi^3_{\Delta}(A,B,C) &=& \Phi^2_{\Delta}(A,BC) - \Phi^2_{\Delta}(A,B)C
- (-1)^{\epsilon_B(\epsilon_A+1)}B\Phi^2_{\Delta}(A,C) \cr
\vdots ~~~ && ~~~\vdots \cr
\Phi^{n+1}_{\Delta}(A_1,\ldots,A_{n+1}) &=& \Phi^n_{\Delta}(A_1,
\ldots,A_nA_{n+1}) - \Phi^n_{\Delta}(A_1,\ldots,A_n)A_{n+1}\cr &&
- (-1)^{\epsilon_{A_{n}}(\epsilon_{A_{1}} + \cdots +\epsilon_{A_{n-1}} +1)}
A_n\Phi^n_{\Delta}(A_1,\ldots,A_{n-1},A_{n+1}) ~.
\end{array}
\label{Phidef2}
\eeq
If $\Phi^k_{\Delta}$ acts like a derivation, $\Phi^{k+1}_{\Delta}$ vanishes
identically, and the iteration terminates.

\noi
When $\Phi^2_{\Delta}$ fails to act like a derivation of the kind
(\ref{Leibniz}), it also fails to fulfill the Jacobi identity (\ref{Jacobi}).
Instead, one finds
\begin{eqnarray}
\sum_{\mbox{\rm cycl.}}(-1)^{(\epsilon_A+1)(\epsilon_C+1)}(A,(B,C))
&=& (-1)^{\epsilon_A(\epsilon_C+1)+\epsilon_B+\epsilon_C}\Phi^1_{\Delta}(
\Phi^3_{\Delta}(A,B,C)) \cr
&&+ \sum_{\mbox{\rm cycl.}}(-1)^{\epsilon_A(\epsilon_C
+1)+\epsilon_B+\epsilon_C}\Phi^3_{\Delta}(\Phi^1_{\Delta}(A),B,C) ~.
\end{eqnarray}
So $\Phi^3_{\Delta}$ equivalently measures the failure of the Jacobi
identity for the usual antibracket.\footnote{This result has a
well-known analogy in the theory of even (Poisson) brackets.} In terms
of the $\Phi^n_{\Delta}$'s themselves, the (broken) Jacobi identity
takes the form
\begin{eqnarray}
\sum_{\mbox{\rm cycl.}}(-1)^{\epsilon_A(\epsilon_C+1)}\Phi^2_{\Delta}
(A,\Phi^2_{\Delta}(B,C))
&=& (-1)^{\epsilon_A\epsilon_C+1}\Phi^1_{\Delta}(
\Phi^3_{\Delta}(A,B,C)) \cr
&&+ \sum_{\mbox{\rm cycl.}}(-1)^{\epsilon_A\epsilon_C
+1}\Phi^3_{\Delta}(\Phi^1_{\Delta}(A),B,C) ~.
\end{eqnarray}

\noi
The above construction shows explicitly that $\Phi^n_{\Delta}$ can be
defined directly in terms of the lowest bracket $\Phi^1_{\Delta}$. However,
the defining equations are highly cumbersome when $n$ is large, and it
is therefore useful to have a more compact formulation. In order to be
more precise, we will introduce some mathematical notation that turns
out to be very convenient. Because we wish to compare directly with
Koszul \cite{Koszul}, we will give up the condition that the
$\Delta$-operator is based on right-derivatives (as is natural from the
BRST-charge definition, and the Batalin-Vilkovisky formalism) and allow
it to act as a higher-order left-derivative (as is more natural from
the mathematical point of view). The translation between the two
conventions is of course trivial. To avoid confusion, the analogous
$\Delta$-operators will in the following be denoted by capital roman
letters $S, T$, etc.

\subsection{\sc An Algebraic Definition}

\noi
Let ${\cal A}$ be a supercommutative algebra with unit  ${\bf 1}$ over
the complex field $C$. Furthermore, let $T{\cal A}$ denote the
tensor algebra of ${\cal A}$:
\beq
  T{\cal A} = \sum_{n=0}^{\infty} {\cal A}^{\otimes n}
    =C + {\cal A} +  {\cal A} \otimes {\cal A}
    +  {\cal A} \otimes {\cal A} \otimes {\cal A} + \ldots~.
\eeq
We distinguish between the unit element in the algebra
${\bf 1} \in {\cal A}$ and the unit element in the field $1 \in C$
by using boldface type for the algebra unit.
Note in particular that $1 \otimes A= 1 \cdot A = A \in {\cal A}$, but
${\bf 1} \otimes A \in {\cal A} \otimes {\cal A}$ for an element
$A \in {\cal A}$.

\noi
The quotient algebra $S{\cal A}=T{\cal A}/I$ is the (super)symmetrized
tensor algebra of ${\cal A}$, where $I$ denotes the two-sided ideal
generated by the (super)commutator, i.e.\ elements of the form:
\beq
\left[ A \stackrel{\otimes}{,} B \right] \equiv
A\otimes B - (-1)^{\epsilon_A\epsilon_B}B\otimes A ~, ~
A,B \in {\cal A}~.
\eeq
We will mainly work in the (super)symmetrized tensor algebra $S{\cal A}$,
which by construction is an associative and supercommutative algebra
with respect to the tensor product $\otimes$:
\beq
A\otimes B = (-1)^{\epsilon_A\epsilon_B}B\otimes A ~.
\eeq

\noi
It would actually be interesting to do the construction
for an associative but non-commutative algebra ${\cal A}$,
and without super-symmetrizing with respect to the tensor
product.
But for the sake of clarity we will for the moment assume
graded commutativity, and we will also (super)symmetrize the tensor product.
Besides, without guidance from physics it is not obvious which of the
many ways of generalizing to the non-commutative case we should choose.
Akman \cite{Akman} has provided a most natural definition, which
turns out to coincide with a certain expression in terms of supercommutators
which we will provide below.

\noi
Define a multiplication map $^\sim$:$S{\cal A} \rightarrow {\cal A}$,
which takes tensor product $\otimes$ into the product ``$\cdot$'' of
the algebra ${\cal A}$:
\bea
\tilde{1} &=& {\bf 1}  \cr
\tilde{A} &=& A    \cr
 (A_1 \otimes \ldots \otimes A_n)^{\sim} &=& A_1 \cdot \ldots \cdot A_n~.
\eea
For each linear operator $T:{\cal A} \rightarrow {\cal A}$
the composed map $T \circ ^\sim:S{\cal A} \rightarrow {\cal A}$ is
also, in a slight abuse of notation, denoted by $T$.
In particular, we point out that with this definition
$T(1) = T({\bf 1})$.

\noi
At this stage define a co-multiplication (cf. \cite{Koszul})
$\lambda:S{\cal A} \rightarrow S{\cal A} \times S{\cal A} $
\bea
\lambda(1) &=& (1,1)  \cr
 \lambda (A_1 \otimes \ldots \otimes A_n)
&=& ((A_1 , 1) - (1 , A_1)) 
\otimes \ldots \otimes  ((A_n , 1) - (1 , A_n)) ~.
\label{Koszullambda}
\eea
Here $S{\cal A} \times S{\cal A}$ is equipped
with a graded product $\otimes$:
\beq
 (A, B) \otimes (C,D) = (-1)^{\epsilon_B \epsilon_C }
(A \otimes C, B \otimes D) ~.
\eeq
We can understand the curious sign-factor as originating from
permuting $B$ and $C$.
$S{\cal A} \times S{\cal A} \cong S{\cal A} \otimes S{\cal A}
{\ \lower-1.2pt\vbox{\hbox{\rlap{$>$}\lower5pt\vbox{\hbox{$\sim$}}}}\ }
S{\cal A}$ has a canonical map onto $S{\cal A}$,
where the cross product $\times$ is substituted with the tensor product
$\otimes$.

\noi
We now define a map $\Phi_T: S{\cal A} \rightarrow S{\cal A}$
for a linear operator $T$ as
\beq
 \Phi_T \equiv (T \times {\rm Id}_{S{\cal A}}) \circ \lambda~.
\label{PhiT}
\eeq
In this way $ \Phi_T(1) = T({\bf 1})$, while $ \Phi_T({\bf 1}) =
T({\bf 1})-T({\bf 1}) \otimes {\bf 1} $. The operator $T$ only operates
on the first copy of $S{\cal A}$ in $S{\cal A} \times S{\cal A}$
while leaving the second copy untouched. We can invoke this action
for practical calculations with the help of an omit operator
${\wedge} ^{T} : S{\cal A} \rightarrow S{\cal A}$
\bea
   \widehat{1}^{T}&=&1 \cr
   (A \otimes B)^{\wedge T} &=& \widehat{A}^{T} \otimes \widehat{B}^{T} \cr
   T(A \otimes \widehat{B}^{T}) &=&  (T(A) , B) \approx T(A) \otimes B ~.
\eea
So whenever an argument of $T$ is decorated with the omit-operator,
the argument should be removed from the argument-list of $T$,
and appear outside to the right (or left) instead.
We emphasize that the omit-operation in general involves a sign factor.
For instance,
\beq
T(\widehat{A}^{T}\otimes B) = (-1)^{\epsilon_A\epsilon_B }T(B)\otimes A ~.
\eeq

\noi
With this definition we can write
\bea
  \lefteqn{  \Phi_T(A_1 \otimes \ldots \otimes A_n) = T((A_1-
\widehat{A_1}^T) \otimes \ldots \otimes (A_n- \widehat{A_n}^T))} \cr
&=& \sum_{i_1, \ldots ,i_n =0}^1 (-1)^{\sum_{j>k}\epsilon_{A_j}
\epsilon_{A_k} i_j (1-i_k) }T(A_1^{i_1} \otimes \ldots \otimes A_n^{i_n})
 \otimes (-A_1)^{1-i_1} \otimes \ldots \otimes (-A_n)^{1-i_n} ~.
\eea
We have here employed the obvious conventions $A^0\equiv 1$ and
$A^1\equiv A$.
A useful way of writing this is
\beq
\Phi_T(A_1\otimes\ldots\otimes A_n) = \left[ \left[ \ldots
\left[\stackrel{\rightarrow}{T}
\stackrel{\otimes}{,}A_1 \right]\stackrel{\otimes}{,} \ldots
\right]\stackrel{\otimes}{,}
A_n\right]1~ ,
\eeq
where $\stackrel{\rightarrow}{T}$ operates on every argument to the right.

\noi
At the present stage the connection between the map $\Phi_T$ and the
corresponding higher antibrackets $\Phi^n_T$ may not yet be obvious.
Roughly, the commas used to separate the entries in the higher antibrackets
in the previous subsection have been replaced by the tensor products here.
This is of course only a matter of notation, and clearly immaterial. (And
we shall freely alternate between the two ways of writing it). To see
that we are really very close to having defined the higher antibrackets
$\Phi^n_T$, let us evaluate the lowest cases of $\Phi_T$:
\bea
\Phi_T(1) &=& T({\bf 1}) \cr
\Phi_T(A) &=&\left[\stackrel{\rightarrow}{T}
\stackrel{\otimes}{,}A \right]1
\ = \  T(A)-T({\bf 1}) \otimes A \cr
\Phi_T(A \otimes B) &=&\left[\left[\stackrel{\rightarrow}{T}
\stackrel{\otimes}{,}A \right]\stackrel{\otimes}{,}B \right]1 \cr
&=&T(A \otimes B ) -T(A) \otimes B
    - (-1)^{\epsilon_A \epsilon_B } T(B) \otimes A
     + T({\bf 1}) \otimes A  \otimes B \cr
\Phi_T(A \otimes B \otimes C) &=&\left[\left[\left[\stackrel{\rightarrow}{T}
\stackrel{\otimes}{,}A \right]\stackrel{\otimes}{,}B \right]
\stackrel{\otimes}{,}C \right]1 \cr
&=&T(A \otimes B \otimes C )
-T(A \otimes B) \otimes C - (-1)^{\epsilon_A (\epsilon_B+\epsilon_C) }
T(B \otimes C) \otimes A \cr
&& -(-1)^{(\epsilon_A+\epsilon_B)\epsilon_C }T(C \otimes A) \otimes B
+T(A) \otimes B \otimes C
+ (-1)^{\epsilon_A (\epsilon_B+\epsilon_C) }T(B) \otimes C \otimes A \cr
&& +(-1)^{(\epsilon_A+\epsilon_B)\epsilon_C }T(C) \otimes A \otimes B
- T({\bf 1}) \otimes A  \otimes B \otimes C ~.
\eea

\noi
The higher antibracket $\Phi^n_T: S^n{\cal A} \rightarrow {\cal A}$
of order $n$ is now finally defined by
\beq
\Phi^n_T \equiv \left. \left. \widetilde{\Phi}_T \right|_{S^n{\cal A}}
\equiv    \widetilde{\Phi}_T \circ \pi \right._{S^n{\cal A}} ~.
\eeq
This means that
\bea
\Phi^n_T(A_1 \otimes \ldots \otimes A_n)
&=& \left( T((A_1- \widehat{A_1}^T) \otimes \ldots
\otimes (A_n- \widehat{A_n}^T)) \right)^{\sim} \cr
&=& \sum_{i_1, \ldots ,i_n =0}^1 (-1)^{\sum_{j>k}\epsilon_{A_j}
\epsilon_{A_k}i_j (1-i_k) }
 T(A_1^{i_1} \ldots A_n^{i_n})
  (-A_1)^{1-i_1}  \ldots  (-A_n)^{1-i_n}  ~.
\label{PhinT}
\eea
We emphazise a particular useful representation of $\Phi^n_T$:
\beq
\Phi^n_T(A_1\otimes\ldots\otimes A_n) = \left[ \left[ \ldots \left[
\stackrel{\rightarrow}{T},A_1 \right],\ldots , \right]
A_n\right] {\bf 1}
\label{comrep} ~.
\eeq
This immediately leads to the following recursion relation:
\bea
\Phi^{n+1}_T(A_1 \otimes \ldots  \otimes A_{n+1})
&=& \Phi^n_T(A_1 \otimes \ldots  \otimes A_nA_{n+1})
- \Phi^n_T(A_1 \otimes \ldots  \otimes A_n)A_{n+1} \cr
&& - (-1)^{\epsilon_{A_{n}} \epsilon_{A_{n+1}}}
\Phi^n_T(A_1 \otimes \ldots  \otimes A_{n-1} \otimes  A_{n+1}) A_n
\label{recur}~.
\eea
which agrees with that of eq. (\ref{Phidef2}).

\noi
Finally, let us evaluate some of the lowest cases:
\beq
\begin{array}{rcccl}
   \Phi^0_T(1) &=& T{\bf 1} &=& T({\bf 1})  \\
  \Phi^1_T(A) &=& \left[T,A \right] {\bf 1}  &=&T(A)-T({\bf 1})  A \\
  \Phi^2_T(A\otimes B) &=&\left[ \left[T,A \right], B \right] {\bf 1}
&=&T(AB) - T(A)B - (-1)^{\epsilon_A\epsilon_B }T(B)A + T({\bf 1})AB~.
\end{array}\label{commutatordef}
\eeq
Specializing to the case of $T({\bf 1}) = T(1) = 0$, this definition is
seen to agree with the one of eq. (\ref{Phidef1}), once translated into
an operator $T$ differentiating from the left. The more general definition
with $T({\bf 1})$ not necessarily vanishing can of course (since the
above considerations are based on Koszul's construction) be found in ref.
\cite{Koszul} as well.

\noi
Normally, $T$ is a differential operator.
Note that if  $T$ is a (left) multiplication operator, then all brackets
vanish identically, except for the zero bracket.

\noi
It may also be of interest to note that it is possible to invert the
relation between the operator $T$ and $\Phi_T$.
One way is to project  $\Phi_T$ into the algebra ${\cal A}$ itself :
\mbox{$\pi_{{\cal A}} \circ \Phi_T = T$}.
The following relations hold in the tensor algebra as well:
\bea
   \Phi_T((A_1+ \widehat{A_1}^{ \Phi_T}) \otimes \ldots
       \otimes (A_n+ \widehat{A_n}^{ \Phi_T}))
    &=&  T(A_1 \otimes \ldots \otimes A_n)
= \widetilde{T}(A_1 \otimes \ldots \otimes A_n)  \cr
  &=&  \left( \widetilde{\Phi}_T((A_1+ \widehat{A_1}^{ \widetilde{\Phi}_T})
\otimes \ldots  \otimes
 (A_n+ \widehat{A_n}^{ \widetilde{\Phi}_T}))   \right)^{\sim} ~.
\label{invPhiT}
\eea

\subsection{\sc The Strongly Homotopy Lie Algebra}

\noi
There is an intriguing connection between the algebra of higher
antibrackets based on Grassmann odd and nilpotent operators,
and strongly homotopy Lie algebras \cite{Stasheff,Lada}.

\noi
{\bf LEMMA:~}
Let \mbox{$S,T \in {\rm Hom}_{C}({\cal A},{\cal A})$} and assume
${\cal A}$ is an algebra (and hence with a product). Then
\beq
 \Phi_{ST} = \Phi_S \circ b_{\textstyle \widetilde{\Phi}_{T} }
+ \left| \Phi_{S}, \widetilde{\Phi}_{T} \right|   
\label{lemma}
\eeq
and  (by operating with tilde on both sides)
\beq
  \widetilde{\Phi}_{ST} = \widetilde{\Phi}_S \circ b_{\textstyle
\widetilde{\Phi}_{T} }
+ \left\{\widetilde{\Phi}_{S}, \widetilde{\Phi}_{T} \right|  
\eeq
Here the co-derivation \mbox{$b_{\textstyle \widetilde{\Phi}_{T}}$} 
is defined as
\beq
  b_{\textstyle \widetilde{\Phi}_{T}}(A_1 \otimes \ldots \otimes A_n)
 = \sum_{i_1, \ldots ,i_n =0}^1 (-1)^{\sum_{j>k}\epsilon_{A_j}
\epsilon_{A_k} i_j (1-i_k) }
\widetilde{\Phi}_{T}(A_1^{i_1} \otimes \ldots \otimes A_n^{i_n})
 \otimes A_1^{1-i_1} \otimes \ldots \otimes A_n^{1-i_n} ~.
\eeq
The Lemma also contain the first example of a {\em bracket-brackets}   
\mbox{$\left| \Phi_{S}, \widetilde{\Phi}_{T} \right|$}:
\bea
  \left.  \left| \Phi_{S}, \widetilde{\Phi}_{T} \right|
A_1 \otimes \ldots \otimes A_n \right\}
&=& \sum_{r=0}^n {1\over {r! (n-r)! }}
\sum_{\pi \in {\cal S}_n }
(-1)^{\epsilon_{\pi}+\epsilon_{T}(\epsilon_{A_{\pi(1)}}
+ \ldots +\epsilon_{A_{\pi(r)}} )}
\Phi_S( A_{\pi(1)} \otimes \ldots \otimes A_{\pi(r)}  )  \cr
&& ~~~~~~~~~~~~~~~~~~~~~~~~~~~~~~~~~~~~ \otimes \
 \Phi^{n-r}_T (A_{\pi(r+1)} \otimes \ldots \otimes A_{\pi(n)} )
 ~.
\eea
This is the simplest of an infinite tower of bracket-brackets. 
One can associate a tilded pendant  
\beq
 \left\{\widetilde{\Phi}_{S}, \widetilde{\Phi}_{T} \right|  
\ = \  \left| \Phi_{S}, \widetilde{\Phi}_{T} \right|^{\sim}  ~.
\eeq
We refer to appendix A and B for a throughout presentation 
of co-derivation and bracket-brackets. 
Here we will merely note that the second term in (\ref{lemma}) 
with these generalizations
can take the following disguises:
\beq
 \left\{\widetilde{\Phi}_{S}, \widetilde{\Phi}_{T} \right|
= \left\{\Phi_{S}, \Phi_{T} \right|
= \left\{\left| \matrix{
S&  {\rm Id}_{S{\cal A}} \cr
1  & -1 \cr
}\right|
, \left| \matrix{
T &  {\rm Id}_{S{\cal A}} \cr
1  & -1 \cr
}\right|
 \right|
= \left\{ \matrix{
S &  {\rm Id}_{S{\cal A}} & T &  {\rm Id}_{S{\cal A}}\cr
1  & -1 & 1 & -1 \cr
}\right|
= \left\{ \matrix{
S  & T &  {\rm Id}_{S{\cal A}}\cr
1   & 1 & -2 \cr
}\right|
\eeq
Let us insert arguments $A_1, \ldots , A_n$.
The lemma can then be stated as
\bea
 \lefteqn{ \Phi_{ST} (A_1 \otimes \ldots \otimes A_n) } \cr
&=& \sum_{r=0}^n {1\over {r! (n-r)! }}
\sum_{\pi \in {\cal S}_n }
(-1)^{\epsilon_{\pi}} \Phi_S\left(
\Phi^r_T (A_{\pi(1)} \otimes \ldots \otimes A_{\pi(r)}  )
 \otimes A_{\pi(r+1)} \otimes \ldots \otimes A_{\pi(n)} \right) \cr
&&+ \sum_{r=0}^n {1\over {r! (n-r)! }}
\sum_{\pi \in {\cal S}_n }
(-1)^{\epsilon_{\pi}+\epsilon_{T}(\epsilon_{A_{\pi(1)}}
+ \ldots +\epsilon_{A_{\pi(r)}} )}
\Phi_S( A_{\pi(1)} \otimes \ldots \otimes A_{\pi(r)}  )  \cr
&& ~~~~~~~~~~~~~~~~~~~~~~~~~~~~~~~~~~~~ \otimes \
 \Phi^{n-r}_T (A_{\pi(r+1)} \otimes \ldots \otimes A_{\pi(n)} )
 ~.
\eea
$\epsilon_{\pi}$ is the Grassmann parity originating
from permuting Grassmann graded quantities:

\beq
   A_{\pi(1)} \dots A_{\pi(n)}
=(-1)^{\epsilon_{\pi} } A_1 \dots A_n
\eeq

\noi
\underline{Proof of lemma:}
It is clearly enough to prove the lemma for bosonic arguments $A_1, \ldots ,
A_n$.
The first term on the righthand side is:
\bea
\lefteqn{  \sum_{r=0}^n {1\over {r! (n-r)! }}
\sum_{\pi \in {\cal S}_n } \Phi_S\left(
\Phi^r_T (A_{\pi(1)} \otimes \ldots \otimes A_{\pi(r)} )
 \otimes A_{\pi(r+1)} \otimes \ldots \otimes A_{\pi(n)} \right) }\cr
&=&\sum_{i_1, \ldots ,i_n =0}^1
\Phi_S \left( \widetilde{\Phi}_T (
 A_1^{i_1} \otimes \ldots \otimes A_n^{i_n} )
 \otimes A_1^{1-i_1} \otimes \ldots \otimes A_n^{1-i_n} \right) \cr
&=&\sum_{i_1, \ldots ,i_n =0}^1   \sum_{j_1=0}^{i_1} \ldots \sum_{j_n=0}^{i_n}
\Phi_S \left( T ( A_1^{i_1 j_1}  \ldots  A_n^{i_n j_n}  )
 (-A_1)^{i_1(1-j_1)} \ldots  (-A_n)^{i_n(1-j_n)}
 \otimes A_1^{1-i_1} \otimes \ldots \otimes A_n^{1-i_n} \right) \cr
&=&\sum_{i_1, \ldots ,i_n =0}^1   \sum_{j_1=0}^{i_1} \ldots \sum_{j_n=0}^{i_n}
\sum_{k_0=0}^1 \sum_{k_1=i_1}^1 \ldots \sum_{k_n=i_n}^1   \cr
&& S \left( \left( T  ( A_1^{i_1 j_1}  \ldots  A_n^{i_n j_n}  )
 (-A_1)^{i_1(1-j_1)} \ldots  (-A_n)^{i_n(1-j_n)} \right)^{k_0}
\otimes A_1^{(1-i_1)(1-k_1)}  \otimes\ldots \otimes A_n^{(1-i_n)(1-k_n)}
\right)
\cr
&& \otimes  \left(- T ( A_1^{i_1 j_1}  \ldots  A_n^{i_n j_n} )
 (-A_1)^{i_1(1-j_1)} \ldots  (-A_n)^{i_n(1-j_n)} \right)^{1-k_0}
 \otimes (-A_1)^{(1-i_1)k_1}  \otimes\ldots\otimes  (-A_n)^{(1-i_n)k_n} \cr
&=&   (k_0=0){\rm -terms}  \  + \  (k_0=1){\rm -terms} ~.
\eea
It is straight forward to see that
the $(k_0=1)$-terms are the left hand side of the lemma:
\bea
(k_0=1){\rm -terms}
&=&\sum_{i_1, \ldots ,i_n =0}^1   \sum_{j_1=0}^{i_1} \ldots \sum_{j_n=0}^{i_n}
 \sum_{k_1=i_1}^1 \ldots \sum_{k_n=i_n}^1
 S \left(  T ( A_1^{i_1 j_1}  \ldots  A_n^{i_n j_n}  )  \right.  \cr
&&~~~~~~~~~~ \left. \times \  (-A_1)^{i_1(1-j_1)} \ldots  (-A_n)^{i_n(1-j_n)}
\otimes \ A_1^{(1-i_1)(1-k_1)} \otimes \ldots\otimes  A_n^{(1-i_n)(1-k_n)}
\right) \cr
&& ~~~~~~~~~~~~~~~~~~~~~~~~~~~~~~~~~~~~\otimes \
 (-A_1)^{(1-i_1)k_1}  \otimes\ldots\otimes  (-A_n)^{(1-i_n)k_n}  \cr && \cr
&=& \sum_{\ell_1, \ldots ,\ell_n =0}^1
 ST(A_1^{\ell_1}  \ldots  A_n^{\ell_n})
\otimes (-A_1)^{1-\ell_1}  \otimes\ldots\otimes  (-A_n)^{1-\ell_n}  \cr
&=& \Phi_{ST} (A_1 \otimes \ldots \otimes A_n) ~,
\eea
due to a cancellation between terms in which
$S$ is not operating directly opon $T$.
Note that in case of  $k_0=0$
the $S$- and $T$-expressions are always multiplied.
The $(k_0=0)$-terms
are minus the second term on the righthand side in the lemma:
\bea
- \  (k_0=0){\rm -terms}
&=&\sum_{i_1, \ldots ,i_n =0}^1   \sum_{j_1=0}^{i_1} \ldots \sum_{j_n=0}^{i_n}
 \sum_{k_1=i_1}^1 \ldots \sum_{k_n=i_n}^1
 S (  A_1^{(1-i_1)(1-k_1)} \otimes \ldots \otimes A_n^{(1-i_n)(1-k_n)}  ) \cr
&&  ~~~~~~~~~~~~~~~~~~~~~~~~~~~~~~~~~~~~ \otimes \
T ( A_1^{i_1 j_1}  \ldots  A_n^{i_n j_n} )
   (-A_1)^{i_1(1-j_1)} \ldots  (-A_n)^{i_n(1-j_n)} \cr
&& ~~~~~~~~~~~~~~~~~~~~~~~~~~~~~~~~~~~~ \otimes \
(-A_1)^{(1-i_1)k_1} \otimes \ldots \otimes (-A_n)^{(1-i_n)k_n}  \cr
&=& \sum_{\ell_1, \ldots ,\ell_n =0}^1
 \Phi_S(A_1^{\ell_1} \otimes\ldots\otimes  A_n^{\ell_n})
\otimes \widetilde{\Phi}_{T}(A_1^{1-\ell_1}  \otimes\ldots\otimes
A_n^{1-\ell_n} ) \cr
&=&  \left.  \left| \Phi_{S}, \widetilde{\Phi}_{T} \right|
A_1 \otimes \ldots \otimes A_n \right\}
 ~.
\eea
\begin{flushright}
${\,\lower0.9pt\vbox{\hrule \hbox{\vrule
height 0.2 cm \hskip 0.2 cm \vrule height 0.2 cm}\hrule}\,}$
\end{flushright}

\noi
An anti-supersymmetrization in $S$ and $T$ of the tilded version of the lemma
cause the second terms to drop out:\footnote{Here, and throughout our
paper, $[A,B]$ denotes the graded commutator: $[A,B] \equiv AB -
(-1)^{\epsilon_A\epsilon_B}BA$.}
\beq
  \widetilde{\Phi}_{[S,T]} = \widetilde{\Phi}_{[S}  \circ b_{\textstyle
\widetilde{\Phi}_{T]} }~,
\label{TSmainid}
\eeq
or equivalently, with arguments $A_1, \ldots , A_n$ inserted:
\bea
&&  \Phi^n_{[S,T]} (A_1 \otimes \ldots \otimes A_n)
= \cr && \sum_{r=0}^n {1\over {r! (n-r)! }}
\sum_{\pi \in {\cal S}_n }
(-1)^{\epsilon_{\pi}} \Phi^{n-r+1}_{[S}\left(
\Phi^r_{T]}  (A_{\pi(1)} \otimes \ldots \otimes A_{\pi(r)}  )
 \otimes A_{\pi(r+1)} \otimes \ldots \otimes A_{\pi(n)} \right)   ~.
\label{mid}
\eea
This contains the main identities for strongly homotopy Lie algebras.
(We borrow the terminology ``main identity'' from closed string field
theory \cite{Zwiebach}, where analogous expressions play an important
r\^{o}le; see section 3).
Let us write out the first few identities.

\noi
\underline{$n=0$:}
\beq
  \Phi^0_{[S,T]}
=  \Phi^1_{[S} ( \Phi^0_{T]} )   ~.
\eeq

\noi
\underline{$n=1$:}
\beq
  \Phi^1_{[S,T]} (A)
=  \Phi^2_{[S} ( \Phi^0_{T]} \otimes A  )
+ \Phi^1_{[S} ( \Phi^1_{T]} \left(  A  \right)  )   ~.
\eeq

\noi
\underline{$n=2$:} ~~{\em Leibnitz rule for a (not necessarily odd)
Laplacian and associated (anti)bracket}
\bea
  \Phi^2_{[S,T]} (A_1 \otimes  A_2)
&=& \Phi^3_{[S}\left( \Phi^0_{T]} \otimes A_1 \otimes  A_2 \right) \cr
&&+ \sum_{\pi \in {\cal S}_2 }
(-1)^{\epsilon_{\pi}} \Phi^2_{[S}\left(
\Phi^1_{T]}  (A_{\pi(1)}   )
 \otimes A_{\pi(2)}  \right)  \cr
&&+  \Phi^1_{[S}\left( \Phi^2_{T]} \left(A_1 \otimes  A_2 \right) \right) \cr
&=& \Phi^3_{[S}\left( \Phi^0_{T]} \otimes A_1 \otimes  A_2 \right) \cr
&&+
 \Phi^2_{[S}\left(
\Phi^1_{T]} \left(A_1  \right)
 \otimes A_2  \right)  \cr
&&+(-1)^{\epsilon_{A_1}\epsilon_{A_2}} \Phi^2_{[S}\left(
\Phi^1_{T]} \left(A_2  \right)
 \otimes A_1  \right)  \cr
&&+  \Phi^1_{[S}\left( \Phi^2_{T]} \left(A_1 \otimes  A_2 \right) \right) ~.
\label{leiblaplanti}
\eea

\noi
\underline{$n=3$:} ~~{\em Jacobi identity}
\bea
 \Phi^3_{[S,T]} (A_1 \otimes A_2 \otimes A_3)
&=&  \Phi^4_{[S}\left(
\Phi^0_{T]} \otimes A_1 \otimes A_2 \otimes A_3 \right)  \cr
&&+ { 1 \over 2} \sum_{\pi \in {\cal S}_3 }
(-1)^{\epsilon_{\pi}} \Phi^3_{[S}\left(
\Phi^1_{T]}  (A_{\pi(1)}  )
 \otimes A_{\pi(2)}  \otimes A_{\pi(3)} \right) \cr
&& +  { 1 \over 2} \sum_{\pi \in {\cal S}_3 }
(-1)^{\epsilon_{\pi}} \Phi^2_{[S}\left(
\Phi^2_{T]}  (A_{\pi(1)} \otimes A_{\pi(2)}  )
 \otimes  A_{\pi(3)} \right) \cr
&&+ \Phi^1_{[S}\left(
\Phi^3_{T]} \left(A_1 \otimes A_2 \otimes A_3 \right) \right) \cr
&=&  \Phi^4_{[S}\left(
\Phi^0_{T]} \otimes A_1 \otimes A_2 \otimes A_3 \right)  \cr
&&+ \Phi^3_{[S}\left(
\Phi^1_{T]} \left(A_1 \right)
 \otimes A_2  \otimes A_3 \right) \cr
&&+(-1)^{\epsilon_{A_1}(\epsilon_{A_2}+\epsilon_{A_3})}
\Phi^3_{[S}\left(
\Phi^1_{T]} \left(A_2 \right)
 \otimes A_3  \otimes A_1 \right) \cr
&&+(-1)^{(\epsilon_{A_1}+\epsilon_{A_2})\epsilon_{A_3}}
\Phi^3_{[S}\left(
\Phi^1_{T]} \left(A_3 \right)
 \otimes A_1  \otimes A_2 \right) \cr
&& + \Phi^2_{[S}\left(
\Phi^2_{T]} \left(A_1 \otimes A_2 \right)
 \otimes  A_3 \right) \cr
&&+(-1)^{\epsilon_{A_1}(\epsilon_{A_2}+\epsilon_{A_3})}
\Phi^2_{[S}\left(
\Phi^2_{T]} \left(A_2 \otimes A_3 \right)
 \otimes  A_1 \right) \cr
&&+(-1)^{(\epsilon_{A_1}+\epsilon_{A_2})\epsilon_{A_3}}
\Phi^2_{[S}\left(
\Phi^2_{T]} \left(A_3 \otimes A_1\right)
 \otimes  A_2 \right) \cr
&&+ \Phi^1_{[S}\left(
\Phi^3_{T]} \left(A_1 \otimes A_2 \otimes A_3 \right) \right)   ~.
\eea
It is quite amazing that the main identities for strongly homotopy Lie
algebras, which in closed string field theory rely on non-trivial
geometric properties in moduli space \cite{Zwiebach}, here can be derived
as a purely
algebraic result due to an assumed existence of a product
(so that ${\cal A}$ is an algebra, and not just a vector space).
If one does not assume the existence of this product, one can reformulate
the right hand side of the main identity (\ref{TSmainid}) in terms of
nilpotency of co-derivations $b_{\textstyle \widetilde{\Phi}_{T} } $:
\beq
\widetilde{\Phi}_{[S}  \circ b_{\textstyle \widetilde{\Phi}_{T]} } =0
 \ \ \Leftrightarrow  \ \
b_{\textstyle \widetilde{\Phi}_{[S} 
\circ b_{\textstyle \widetilde{\Phi}_{T]} }
} =0
\ \ \Leftrightarrow  \ \
b_{\textstyle \widetilde{\Phi}_{[S}} 
\circ b_{\textstyle \widetilde{\Phi}_{T]}
}
=0
\eeq
This follows quite easily from (\ref{bsim}) and (\ref{bder}).

\noi
\subsection{\sc  Coordinate Representation}

\noi
We will now translate the above construction into a description
with explicitly chosen coordinates.
Let \mbox{$\left\{\left. e_a \right| a \in I \right\}$} denote a vector basis
for ${\cal A}$,
and \mbox{$\left\{\left. \eta^a \right| a \in I \right\}$}
the dual basis in  ${\cal A}^{\ast}$, so that
\beq
   \eta^a ( e_b) = \delta^a_b
\eeq
Without loss of generality we can take
the coordinates $A^a$ of a  general element \mbox{$A=\sum_a A^a e_a$}
to be bosonic, $i.e$. the basis vectors are supposed to carry the
Grassmann grading.
Purchasing further the vector space structure of  ${\cal A}$,
one can identify
the space \mbox{${\rm Hom}_{\cal C}(S^n{\cal A}, S^m{\cal A})$}
of linear operators $:S^n{\cal A} \rightarrow S^m{\cal A}$,
with \mbox{$S^m{\cal A}\otimes S^n({\cal A}^{\ast})$},
the set of $S^m{\cal A}$-valued homogeneous polynomials in ${\cal A}$ of
degree $n$:
\beq
  T^{(m,n)} =
e_{a_1}  \ldots  e_{a_m}  \ T^{a_1 \ldots a_m}_{b_1 \ldots b_n}  \
   \eta^{b_1} \ldots \eta^{b_n}~.
\label{Texpansion}
\eeq
Here
\bea
e_{a_1}  \ldots  e_{a_m}
&\equiv& e_{a_1} \otimes \ldots \otimes e_{a_m} \in S^m{\cal A}
\cr && \cr
\eta^{b_1} \ldots \eta^{b_n}
& \in& S^n({\cal A} )^{\ast}  ~ \cong ~ S^n({\cal A}^{\ast} )
\cr && \cr
\eta^{b_1} \ldots \eta^{b_n} \left( e_{a_1}  \ldots  e_{a_m}  \right)
&\equiv& \left\{\begin{array}{l}
 (-1)^{\epsilon_{b}} \sum_{\pi \in {\cal S}_m }
(-1)^{\epsilon_{\pi} }
\delta_{a_{\pi(1)}}^{b_1}  \ldots \delta_{a_{\pi(m)}}^{b_m}
{}~~ {\rm for}~~ n=m \cr \cr
0 ~~ {\rm otherwise}~, \cr
\end{array} \right.
\label{contraction}
\eea
and $\epsilon_{\pi}$ is the Grassmann parity originating
from permuting the Grassmann-graded quantities:
\beq
   e_{a_{\pi(1)}} \dots e_{a_{\pi(n)}}
=(-1)^{\epsilon_{\pi} } e_{a_1} \dots e_{a_n}~.
\eeq
\beq
   \epsilon_{a} ~ \equiv ~
\sum_{i>j}\epsilon_{a_{i}}\epsilon_{a_{j}} ~ ({\rm mod} ~2)~.
\eeq

\noi
To avoid the sign-factor  $\epsilon_{b}$ appearing in (\ref{contraction}),
it is convenient to define a contraction symbol which first organizes all
basis vectors $e_a$  to the right and all
dual vectors $\eta^a$ to the left, and then contracts:
\beq
   \left[ \eta^{a_1} e_{b_1} \ldots \eta^{a_n} e_{b_n}  \right]
{}~~\equiv~~
(-1)^{\epsilon_{a} }
\eta^{a_1}\ldots \eta^{a_n} \left(e_{b_1} \ldots e_{b_n}\right)
{}~~=~~   \sum_{\pi \in {\cal S}_n }
(-1)^{\epsilon_{\pi}}
\delta_{a_{\pi(1)}}^{b_1}  \ldots \delta_{a_{\pi(n)}}^{b_n}~.
\eeq
In other words, the objects (super)commute freely under this contraction
symbol. For fixed set of vectors  \mbox{$e_{a_1}, \ldots, e_{a_m}$},
note that the norm  of a contraction is (no sum over $a_1, \ldots, a_m$):
\beq
\left[ \eta^{a_1} e_{a_1} \ldots \eta^{a_m} e_{a_m}  \right] ~=~
\left\{ \begin{array}{l}
m_1! \ldots m_r ! \cr
0~,
\end{array}
\right. \label{multi}
\eeq
where \mbox{$m_1, \ldots, m_r$} are the multiplicities
of the vectors in the set
 \mbox{$\left\{e_{a_1}, \ldots, e_{a_m}\right\}$}.
\mbox{($m_1+\ldots+m_r=m$)}.
The second alternative in eq. (\ref{multi}) simply occurs when
Grassmann odd vectors have multiplicity $>1$.
If all vectors are odd the norm is therefore either $0$ or $1$.

\noi
Next define a ``symmetrizer projection operator'' by
\beq
P^{a_1 \ldots a_n}_{b_1 \ldots b_n} ~~\equiv~~
 {1\over{  n!}} \sum_{\pi \in {\cal S}_n }
(-1)^{\epsilon_{\pi}}
\delta_{a_{\pi(1)}}^{b_1}  \ldots \delta_{a_{\pi(n)}}^{b_n}
 ~~=~~{1\over{  n!}}
 \left[ \eta^{a_1} e_{b_1} \ldots \eta^{a_n} e_{b_n}  \right] ~.
\eeq
\beq
\begin{array}{rcccccl}
P^{a_1 \ldots a_n}_{b_1 \ldots b_n} P^{b_1 \ldots b_n}_{c_1 \ldots c_n}
{}~~&=&~~P^{a_1 \ldots a_n}_{c_1 \ldots c_n}~,&
{}~~~~~~~~~~~~~~&
(-1)^{\epsilon_{a} } \ P^{a_1 \ldots a_n}_{b_1 \ldots b_n}
{}~&=&~ (-1)^{\epsilon_{b} } \
P^{a_1 \ldots a_n}_{b_1 \ldots b_n} ~, \cr \cr
P^{a_1 \ldots a_n}_{b_1 \ldots b_n}
e_{a_1}  \ldots  e_{a_1} ~&=&~ e_{b_1}  \ldots  e_{b_n} ~,&
{}~~~~~~~~~~~~~~&
P^{a_1 \ldots a_n}_{b_1 \ldots b_n}
\eta^{b_1} \ldots \eta^{b_n}~&=&~ \eta^{a_1} \ldots \eta^{a_n}~.
\end{array}
\eeq
Define the (super)symmetrized coefficients of an operator $T$ by
\bea
 \left(T^{\rm sym} \right)^{a_1 \ldots a_m}_{b_1 \ldots b_n}
{}~&\equiv&~   P^{a_1 \ldots a_m}_{c_1 \ldots c_m} \
T^{c_1 \ldots c_m}_{d_1 \ldots d_n} \
P^{d_1 \ldots d_n}_{b_1 \ldots b_n} \cr &&\cr
{}~&=&~ {{ (-1)^{\epsilon_{a} +  \epsilon_{b} } } \over { n! \ m! }} \
\eta^{a_1} \ldots \eta^{a_m} \left( T ( e_{b_1}  \ldots  e_{b_n} ) \right)~.
\eea
In case of symmetric coefficients this yields an inversion of
eq. (\ref{Texpansion}):
\beq
\left(T^{\rm sym} \right)^{a_1 \ldots a_m}_{b_1 \ldots b_n}
=T^{a_1 \ldots a_m}_{b_1 \ldots b_n} ~.
\eeq

\noi
The composition of two operators
\mbox{$S, T  \in {\rm Hom}_{C}(S{\cal A},S{\cal A})$}~  is then
\beq
    S \circ T ~=~\sum_{k,\ell,m,n=0}^{\infty}
e_{a_1}  \ldots  e_{a_k}  \ S^{a_1 \ldots a_k}_{b_1 \ldots b_{\ell}}  \
  \left( \eta^{b_1} \ldots \eta^{b_{\ell}} \left(e_{c_1}  \ldots  e_{c_m}
\right) \right)
\ T^{c_1 \ldots c_m}_{d_1 \ldots d_n}  \
   \eta^{d_1} \ldots \eta^{d_n}~,
\eeq
or, in terms of coefficients,
\beq
\left((S \circ T)^{\rm sym} \right)^{a_1 \ldots a_m}_{b_1 \ldots b_n}
= \sum_{\ell =0}^{\infty}
 (-1)^{\epsilon_{c} }  \ell !  \
\left(S^{\rm sym} \right)^{a_1 \ldots a_m}_{c_1 \ldots c_{\ell}}
 \left(T^{\rm sym} \right)^{c_1 \ldots c_{\ell}}_{b_1 \ldots b_n} ~.
\eeq

\noi
Let us now define a normal ordering in which
all basis vectors $e_a$ are moved to the left and all
dual vectors $\eta^a$ are moved to the right, while respecting the Grassmann
grading:
\beq
  : e_a \eta^b : ~~\equiv~~ e_a \eta^b ~~~~~,~~~~~ : \eta^a e_b : ~~\equiv~~
(-1)^{\epsilon_{a}\epsilon_{b}}e_b\eta^a ~.
\eeq
We can then write
\bea
  {\rm Id}_{S{\cal A}} &=& : \exp(e_a \eta^a):
{}~=\sum_{k=0}^{\infty}   {{1}\over {k !}}
:e_{a_1} \eta^{a_1} \ldots e_{a_k} \eta^{a_k}:
{}~=\sum_{k=0}^{\infty}  {{ (-1)^{\epsilon_{a} }} \over {k !}}  \
e_{a_1}   \ldots e_{a_k} \  \eta^{a_1} \ldots   \eta^{a_k} ~,
\eea
and co-derivation (cf.\ eq.\ (\ref{bTdef1}-\ref{bTdef2}))
\bea
 b_T&=& :T\exp(e_a \eta^a):
{}~~=~\sum_{k,n=0}^{\infty}{{1}\over{k!}}:e_{a}\ T^{a}_{a_1 \ldots a_n}  \
\eta^{a_1} \ldots \eta^{a_n} \ e_{b_1} \eta^{b_1} \ldots e_{b_k} \eta^{b_k}:
{}~~~.
\eea
Note that the particular bracket \mbox{$\left| T_1 , \ldots , T_k  \right| $}
defined in eq.\ (\ref{specdef}) is just the normal-ordered product:
\beq
\left| T_1 , \ldots , T_k  \right| ~~= ~~: T_1  \ldots  T_k : ~~~~~.
\eeq

\noi
We can represent the dual basis vectors $\eta^a$ by a
left derivative acting to the right:
\beq
   \eta^a~=~ \frac{\rdl}{\delta e_{a}} ~.
\eeq
or analogously represent the basis vectors $e_a$ by a
{\em right} derivative acting to the {\em left}:
\beq
   e_a~=~ \frac{\ldr}{\delta \eta^{a}} ~.
\eeq
Then the contraction (\ref{contraction}) can be written
\beq
\eta^{b_1} \ldots \eta^{b_n} \left( e_{a_1}  \ldots  e_{a_m}  \right)
=  \left[\frac{\rdl}{\delta e_{b_1}}
\ldots  \frac{\rdl}{\delta e_{b_n}},
e_{a_1}  \ldots  e_{a_m}  \right]_{e=0}
=  \left[ \eta^{b_1} \ldots \eta^{b_n} ,
 \frac{\ldr}{\delta \eta^{a_1}} \ldots
 \frac{\ldr}{\delta \eta^{a_m}} \right]_{\eta=0}~.
\eeq
The  conditions $e=0$ resp.\ $\eta=0$ simply ensure
that  the contraction is non-zero only when $n=m$.
Let us at this point mention a handy representation of the symmetrizer
projection operator:
\beq
P^{a_1 \ldots a_m}_{b_1 \ldots b_n} ~=~  \frac{1}{n!}
\frac{\rdr}{\delta \eta^{b_1}}
\ldots  \frac{\rdr}{\delta \eta^{b_n}}
 \left( \eta^{a_1}  \ldots  \eta^{a_m}  \right)_{\eta=0}
\eeq

\noi
An operator  \mbox{$T  \in {\rm Hom}_{C}(S{\cal A},{\cal A})$}
 with precisely one outgoing slot/entry can be represented
by a  vector field operating to the {\em left}:
\beq
\stackrel{\leftarrow}{T}~~=~~ \sum_{n=0}^{\infty}
 \frac{\ldr}{\delta \eta^{a}} \ T^{a}_{a_1 \ldots a_n}  \
   \eta^{a_1} \ldots \eta^{a_n} ~.
\eeq
Note also that the action of $\circ b_{T} $
can be described by the vector field {\em without} letting $\eta=0$:
\bea
S \circ b_{T}  &=&\left[S, \stackrel{\leftarrow}{T}  \right]
= \sum_{k,\ell,n=0}^{\infty}
\frac{\ldr}{\delta \eta^{b_1}} \ldots
\frac{\ldr}{\delta \eta^{b_k}}
\ S^{b_1 \ldots b_k}_{c_1 \ldots c_{\ell}}  \
  \left[ \eta^{c_1} \ldots \eta^{c_{\ell}} ,
\frac{\ldr}{\delta \eta^{a}} \right]
 \ T^{a}_{a_1 \ldots a_n}  \
   \eta^{a_1} \ldots \eta^{a_n}  \cr
&=&\sum_{k,\ell,n=0}^{\infty}
\frac{\ldr}{\delta \eta^{b_1}} \ldots
\frac{\ldr}{\delta \eta^{b_k}} \ \ell
\ \left(S^{\rm sym} \right)^{b_1 \ldots b_k}_{c_1 \ldots c_{\ell -1} a}  \
  \eta^{c_1} \ldots \eta^{c_{\ell -1}}
 \ T^{a}_{a_1 \ldots a_n}  \
   \eta^{a_1} \ldots \eta^{a_n} ~.
\eea
Or, in terms of coordinates,
\beq
\left( S \circ b_{T} \right)^{a_1 \ldots a_m}_{b_1 \ldots b_n}
= \sum_{r=1}^{n}
(-1)^{(\epsilon_{T}+\epsilon_{c}+\epsilon_{b_{r}}+ \ldots + \epsilon_{b_{n}})
(\epsilon_{b_{1}}+ \ldots + \epsilon_{b_{r -1}}) }  \ r
\ \left(S^{\rm sym} \right)^{a_1 \ldots a_m}_{b_{1} \ldots b_{r -1} c}
 \ T^{c}_{b_{r} \ldots b_{n}}  ~.
\eeq
and
\bea
\left(\left( S \circ b_{T} \right)^{\rm sym}
\right)^{a_1 \ldots a_m}_{b_1 \ldots b_n}
&=& \sum_{r=1}^{n}  { 1 \over {n !}} \sum_{\pi \in {\cal S}_n }
(-1)^{(\epsilon_{T}+\epsilon_{c}+\epsilon_{b_{\pi(r)}}+ \ldots +
\epsilon_{b_{\pi(n)}})
(\epsilon_{b_{\pi(1)}}+ \ldots + \epsilon_{b_{\pi(r-1)}}) }  \cr
&& ~~~~~~~~~~~~~~~~
 r \ \left(S^{\rm sym} \right)^{a_1 \ldots a_m}_{b_{\pi(1)} \ldots
b_{\pi(r-1)} c}
 \ T^{c}_{b_{\pi(r)} \ldots b_{\pi(n)}}  ~.
\eea
This has as one important implication that
(the generalized version of) the main identity (\ref{TSmainid}) for a
strongly homotopy Lie algebra
can be formulated as a contraction between vector fields:
\beq
\widetilde{\Phi}_{[S,T]} =\widetilde{\Phi}_{[S}  \circ b_{\textstyle
\widetilde{\Phi}_{T]} }
=\left[ \stackrel{\leftarrow}{{\textstyle \widetilde{\Phi}} }_{[S} ,
\stackrel{\leftarrow}{{\textstyle \widetilde{\Phi}} }_{T]}  \right] ~.
\eeq
In the last expression the larger outer square brackets
denote a contraction {\em i.e}.\ action of the last vector field on
the former, and the smaller inner square brackets means
anti(super)symmetrization in $S$ and $T$.

\noi
The vector field is
\beq
\stackrel{\leftarrow}{{\textstyle \widetilde{\Phi}} }_{T}
{}~~=~~ \sum_{n=0}^{\infty}
 \frac{\ldr}{\delta \eta^{a}}\left.  \Phi^{n}_{T} \right.^{a}_{a_1
\ldots a_n} \eta^{a_1} \ldots \eta^{a_n} ~.
\eeq
and \mbox{$\left. \Phi^{2}_{T} \right.^c_{ab}$} are usual Lie algebra
structure constants. In particular, when $S = T$ and $T^2 = 0$, the whole
main identity of strongly homotopy Lie algebras can then be expressed as
the nilpotency condition of this new vector field. A description of
strongly homotopy Lie algebras in similar terms has been discussed in
ref. \cite{Zwiebach1}. Stasheff \cite{Stasheff}
expresses the main identity of strongly homotopy Lie algebras in an
analogous way, but without going to particular coordinates.

\noi
Notice that the main identity takes the following form in terms of
symmetrized components:
\bea
\left(\left. \Phi^{n}_{[S,T]} \right.^{\rm sym}
\right)^{a}_{b_1 \ldots b_n}
&=& \sum_{r=1}^{n}  { 1 \over {n !}} \sum_{\pi \in {\cal S}_n }
(-1)^{(\epsilon_{T}+\epsilon_{c}+\epsilon_{b_{\pi(r)}}+ \ldots +
\epsilon_{b_{\pi(n)}}) (\epsilon_{b_{\pi(1)}}+ \ldots +
\epsilon_{b_{\pi(r-1)}}) }  \cr
&& ~~~~~~~~~~~~~~~~
  \ r \left. \Phi^{r}_{[S} \right.^{a}_{b_{\pi(1)} \ldots b_{\pi(r-1)} c}
 \left.  \Phi^{n-r+1}_{T]} \right.^{c}_{b_{\pi(r)} \ldots b_{\pi(n)}}
\cr &&\cr
&=& \sum_{r=1}^{n}  { 1 \over {n !}} \sum_{\pi \in {\cal S}_n }
(-1)^{(\epsilon_{T}+\epsilon_{c}+\epsilon_{b_{\pi(n-r+1)}}+ \ldots +
\epsilon_{b_{\pi(n)}})(\epsilon_{b_{\pi(1)}}+ \ldots +
\epsilon_{b_{\pi(n-r)}}) }  \cr
&& ~~~~~~~~~~~~~~~~
  \ (n-r+1) \left. \Phi^{n-r+1}_{[S} \right.^{a}_{b_{\pi(1)}
\ldots b_{\pi(n-r)} c}   \left.  \Phi^{r}_{T]} \right.^{c}_{b_{\pi(n-r+1)}
\ldots b_{\pi(n)}}   \cr &&\cr &=&  \sum_{r=1}^{n}\frac{1}{n!}
\sum_{\pi \in {\cal S}_n }(-1)^{(\epsilon_{S}+\epsilon_{c}+
\epsilon_{b_{\pi(r)}}+ \ldots + \epsilon_{b_{\pi(n)}})(\epsilon_{T}+
\epsilon_{c}+ \epsilon_{a})  }  \cr && ~~~~~~~~~~~~~~~~
  \ r  \left.  \Phi^{n-r+1}_{[S} \right.^{c}_{b_{\pi(r)} \ldots b_{\pi(n)}}
 \left. \Phi^{r}_{T]} \right.^{a}_{b_{\pi(1)} \ldots b_{\pi(r-1)} c}  ~.
\eea
When written in this form, one also sees that the notion of strongly homotopy
Lie algebras is open to a very natural generalization.

\subsection{\sc A Master Equation and the BRST Symmetry}

\noi
So far all properties of the higher brackets have been derived in a general
frame without any particular applications in mind. 
Clearly, for the usual Batalin-Vilkovisky
Lagrangian quantization program, only one-brackets and two-brackets are
required. This is because the BRST Ward Identities one wishes to impose
on the Lagrangian path integral are Schwinger-Dyson equations. The BRST
operator of Schwinger-Dyson equations can, for flat functional measures,
be chosen to be Abelian \cite{AD3}, and the associated $\Delta$-operator
is then, as explained in section 2, of 2nd order in the appropriate
representation of fields and antifields. But even in the conventional
Lagrangian path integral one may wish to impose other BRST Ward Identities
(subsets of the full set of Schwinger-Dyson equations), and the associated
$\Delta$-operator may then be of higher order \cite{AD1,AD2}. Interestingly,
this imposes the formalism of higher antibrackets as the natural
generalization of the Batalin-Vilkovisky scheme. Both the (quantum)
Master Equation and the (quantum) BRST operator of the Batalin-Vilkovisky
antifield quantization are then seen as very special cases in a much more
general framework. We begin the discussion of this with a few useful
relations.

\noi
We have already seen how the higher brackets can be given a nice
formulation in terms of commutators (see eq. (\ref{commutatordef}).
Let us for later convenience define
a modified operator \mbox{${X}_{T;B_1,\ldots,B_k}$} associated with
the operator $T$:
\beq
\stackrel{\rightarrow}{X}_{T;B_1,\ldots,B_k} (A)
=  \left[ \left[ \ldots \left[\stackrel{\rightarrow}{T}
,B_1 \right], \ldots  \right],B_k\right] A~,
\eeq
where \mbox{$B_1, \ldots , B_k \in {\cal A}$} are fixed elements.
It then follows immediately that
\bea
{X}_{T;A_1,\ldots,A_n} (1) & = & \Phi^{n}_T(
A_1 \otimes \ldots \otimes A_n ) \cr
{X}_{{\textstyle {X}_{T;A_1,\ldots,A_n}};B_1,\ldots,B_k} &=&
{X}_{T;A_1,\ldots,A_n,B_1,\ldots,B_k} ~.\label{composition}
\eea
Notice that this last relation tells us how we can generate higher and
higher brackets by composition!

\noi
Consider the formal exponential function
\beq
e^{\otimes A} = \sum_{n=0}^{\infty} {1 \over {n!}} A^{\otimes n}
= 1 + A + {1 \over 2} A \otimes A
+ {1 \over 6} A \otimes A \otimes A+\ldots \in S{\cal A}  ~.
\eeq
Using this notation, we can write down a very useful formula
\bea
 \widetilde{\Phi}_T(  e^{\otimes A} \otimes B_1 \otimes \ldots \otimes B_k )
&=& \sum_{n=0}^{\infty} {1 \over {n!}}
\Phi^{n+k}_T(  A^{\otimes n}\otimes B_1 \otimes \ldots \otimes B_k ) \cr
&=& e^{-A} \left[ \left[ \ldots \left[\stackrel{\rightarrow}{T}
,B_1 \right], \ldots  \right],B_k\right] e^A \cr
&=& e^{-A} \stackrel{\rightarrow}{X}_{T;B_1,\ldots,B_k} e^A ~.\label{useful}
\eea
The case $k=0$ is just what we would call the quantum Master Equation
\beq
\Delta \exp\left(\frac{i}{\hbar}S \right)= 0
\label{ME}
\eeq
associated with the operator $T$
\beq
 \widetilde{\Phi}_T(  e^{\otimes A} )
= \sum_{n=0}^{\infty} {1 \over {n!}} \Phi^{n}_T(  A^{\otimes n} )
= e^{-A}\stackrel{\rightarrow}{T}e^A ~.
\label{meqtilde}
\eeq
Here $A=\frac{i}{\hbar}S$ is identified with the action,\footnote{Of course
taken to be Grassmann-even.}
and $T$ with the nilpotent Grassmann odd Laplacian:
\beq
T (F) = (-1)^{\epsilon_F} \Delta(F)~.
\eeq
Using the formalism described above, this equation is
easily rewritten in terms of the higher antibrackets
\beq
0 ~=~ \sum_{n=0}^{\infty}\frac{1}{n!}
\left(\frac{i}{\hbar}\right)^n
\Phi^n_{T}(S\otimes \ldots\otimes S)
=e^{- {i \over \hbar} S}\stackrel{\rightarrow}{T}e^{{i \over \hbar} S}
\equiv M(S) ~.
\label{ME1}
\eeq
The name quantum Master Equation is justified by the fact that in the
special case of the Abelian (and 2nd order) Schwinger-Dyson BRST operator
$\Delta$ it reduces to the Batalin-Vilkovisky quantum Master Equation.
Moreover, for more general nilpotent $\Delta$'s it corresponds to the
quantum Master Equation 
when requiring given subsets of this full set of equations (see ref.
\cite{AD2}).

\noi
For given  \mbox{$B_1, \ldots , B_k \in {\cal A}$},
the bracket $\Phi^n_{T}$ (with $n \geq k$)
automatically generates an $(n-k)$-bracket:
\beq
\bar{\Phi}^{n-k}(A_{1} \otimes \ldots \otimes A_{n-k}) \equiv
\Phi^n_{T}(B_1 \otimes \ldots \otimes B_k \otimes A_{1} 
\otimes \ldots \otimes A_{n-k} )
=\Phi^n_{{\textstyle {X}_{T;B_1,\ldots,B_k}}} (A_{1} \otimes \ldots \otimes
A_{n-k})~.
\eeq
In particular, a conventional ``two-antibracket'' $(A,B)$
can always be generated
from the higher antibrackets.
Also, the Master Equation (\ref{ME}) can in
this terminology be seen as the sum  of ``zero-antibrackets''
generated by the action $S$ itself:
\beq
\sum_{k=0}^{\infty}\frac{1}{k!} \left(\frac{i}{\hbar}\right)^k
\Phi^0_{{\textstyle {X}_{T;S,\ldots,S}}} = 0 ~.
\eeq

\noi
Suppose the Master Equation terminates after a finite order of terms, as
happens when $\Delta$ is of finite order:
\beq
\sum_{k=0}^N \left(\frac{i}{\hbar}\right)^k\frac{1}{k!}
\Phi^k_{T}(S\otimes \ldots \otimes S) = 0 ~.
\eeq
{}From the physics perspective it is more natural to view this as an
expansion in $\hbar$:
\beq
\sum_{k=0}^{N}\left(\frac{\hbar}{i}\right)^k \frac{N!}{(N-k)!}
\Phi_{T}^{(N-k)}(S\otimes \ldots \otimes S) = 0 ~.
\eeq
This also suggests a solution $S$ expressed as an $\hbar$-expansion,
beginning with the ``classical action'' $S_0$:
\beq
S ~=~ S_0 + \sum_{n=1}^{\infty}\hbar^n S_n ~.
\eeq
To leading order in the expansion, this leads to the $N$-th order
``classical Master Equation'',
\beq
\Phi_{T}^N(S_0\otimes \ldots \otimes S_0) ~=~ 0 ~,
\eeq
while to next order in $\hbar$ we get
\beq
\Phi_{T}^{N-1}(S_0\otimes \ldots \otimes S_0)
+ i\Phi_{T}^N(S_1\otimes S_0\otimes \ldots \otimes S_0)
= 0 ~,
\eeq
and so on.

\noi
It is curious to note that when the $\Delta$-operator is of infinite order,
and the full Master Equation therefore does not truncate, this solution
in terms of an $\hbar$-expansion loses its meaning. The ``classical''
antibracket is then pushed to infinity, and the analysis must start with
the lowest antibracket $\Delta$ instead.

\noi
In conventional Batalin-Vilkovisky quantization, the BRST operator
is composed of two pieces, a classical part and a ``quantum correction''
(see, $e.g.$, ref. \cite{Henneaux}):
\beq
\sigma^r F ~=~ (F,S) - i\hbar\Delta F ~.
\eeq
We have given $\sigma$ the superscript ``$r$'' to indicate that it acts
with right-derivatives in our conventions (due to $\Delta$).
The most obvious generalisation to the case where the three-brackets (and
perhaps higher brackets as well) do not vanish,
would be (rescaling with a factor ${i \over \hbar}$, and converting to
left derivatives):
\beq
\sigma(F) \equiv \sum_{n=0}^{\infty} {1 \over {n!}}  \left({i \over \hbar}
\right)^{ n}
\Phi^{n+1}_T \left(  F \otimes S^{\otimes n} \right)
= \widetilde{\Phi}_T(  F \otimes e^{\otimes {i \over \hbar} S} )
= e^{-{i \over \hbar} S}  \left[\stackrel{\rightarrow}{T}
,F \right]e^{{i \over \hbar} S} ~.
\label{defsigma}
\eeq
$\sigma$ can be given a meaning purely in terms of higher antibrackets.
The nilpotency of  \mbox{$\sigma$} depends
on the right to use the Master Equation $M(S) = 0$
before all differentiations are
carried out (recall that the brackets in general contains differential
operators):
\beq
  \sigma(\sigma(\epsilon))
 =\widetilde{\Phi}_{T} \left( \widetilde{\Phi}_T(
 \epsilon \otimes e^{\otimes {i \over \hbar} S} ) \otimes
e^{\otimes \frac{i}{\hbar}S} \right)
 =-\widetilde{\Phi}_{T} \left( M(S) \otimes \epsilon \otimes
e^{\otimes \frac{i}{\hbar}S} \right) = 0 ~.
\eeq
The last equality is a consequence of the main identity (\ref{TSmainid}):
\bea
0 &=& \widetilde{\Phi}_{T}  \circ b_{\textstyle \widetilde{\Phi}_{T} }
\left(\epsilon  \otimes e^{\otimes \frac{i}{\hbar}S} \right) \cr
&=&\widetilde{\Phi}_{T} \left( \widetilde{\Phi}_T(
 \epsilon \otimes e^{\otimes {i \over \hbar} S} ) \otimes
e^{\otimes \frac{i}{\hbar}S} \right)
 + \widetilde{\Phi}_{T} \left( \widetilde{\Phi}_T(
 e^{\otimes {i \over \hbar} S} ) \otimes \epsilon \otimes
e^{\otimes \frac{i}{\hbar}S} \right)~.
\eea
Let us test this generalization  \mbox{$\sigma$} by searching for variations
$\delta S$ of the action $S$ that preserve the Master Equation.
Variation \mbox{${i \over \hbar} \delta S=\sigma(\epsilon)$} of the form
(\ref{defsigma})
only preserve the Master Equation (\ref{ME}) ``on-shell"
(where we apply the Master Equation $M(S) = 0$
before all differentiations has been done; this terminology becomes
particularly obvious in the case of string field theory -- see later):
\beq
 \delta \left( \stackrel{\rightarrow}{T}  e^{{i \over \hbar} S}  \right)
=  {i \over \hbar}  \stackrel{\rightarrow}{T}\left(e^{{i \over \hbar} S}
\delta S \right)
\eeq
It may look as if one can have an ``off-shell" invariance of the
Master Equation with respect to a slightly different type of variations:
\beq
 {i \over \hbar}  \delta S ~=~ \bar{\sigma}(\epsilon)
{}~\equiv~ e^{-{i \over \hbar} S} \stackrel{\rightarrow}{T} \left(\epsilon \
e^{{i \over \hbar} S}\right)
\label{defbarsigma}
\eeq
This is however not quite true.
For instance, if one varies the equivalent form (\ref{ME1})
of the master equation, one gets:
\beq
\delta M(S) = {i \over \hbar} \left(- M(S) \delta S
+ e^{-{i \over \hbar} S} \stackrel{\rightarrow}{T}
e^{{i \over \hbar} S} \delta S\right)~.
\eeq
Here variation $\bar{\sigma}(\epsilon)$ of the form (\ref{defbarsigma})
only preserve  (\ref{ME1}) ``on-shell".
However the alternative \mbox{$\bar{\sigma}$} does have the nice property
that nilpotency, \mbox{$\bar{\sigma}^2 = 0$},
is a direct consequence of $T$ being nilpotent.

\noi
The reason why the meaning of ``on-shell" and ``off-shell"
here becomes somewhat obscured, can be traced back to the fact that
neither $\bar{\sigma}$ nor $\sigma$ are derivations,
$i.e.$ do not fulfill the Leibnitz rule.

\noi
Finally, let us mention that in the case of  \mbox{$\sigma$},
the invariance of the master equation (\ref{ME1})
can be directly related to the nilpotency of \mbox{$\sigma$}:
\beq
\delta_{\epsilon} M(S)
={i \over \hbar} \widetilde{\Phi}_{T} \left( \delta_{\epsilon} S \otimes
e^{\otimes \frac{i}{\hbar}S} \right)
 =  \sigma(\sigma(\epsilon))~.
\eeq
Both $\sigma$ and $\bar{\sigma}$ can obviously be viewed as BRST symmetry
operators, and, since $S$ in the BRST context is taken to satisfy the
quantum Master Equation, in fact coincide. From the BRST viewpoint the
fact that deformations $S \to S + \delta S$ of a solution $S$ to the Master
Equation still satisfy this Master Equation is seen as the possibility
of adding BRST-exact terms $\sigma(\epsilon)$ (or $\bar{\sigma}(\epsilon)$)
to the action.

\subsection{\sc The Transformation Algebra}

\noi
When the BRST transformations alternatively are viewed as transformations of
the action $S$, one would like to find the possible
algebra of such transformations. This has already been done in the
framework of the conventional Batalin-Vilkovisky formalism by Hata and
Zwiebach \cite{covariant}
(note that their odd Laplacian consists of
{\em left} derivatives, so we denote it by $T$, to be consistent). Letting
\beq
{i \over \hbar} \delta_{\epsilon} S = \sigma(\epsilon)=
 \stackrel{\rightarrow}{T}   \epsilon +{i \over \hbar} (S , \epsilon)~,
\eeq
they find
\beq
[\delta_{\epsilon_{1}},\delta_{\epsilon_{2}}] F(S)=
 \delta_{(\epsilon_{1},\epsilon_{2})} F(S)~,
\label{2closed}
\eeq
$i.e.$, the algebra of  transformations  on $S$ is just the
algebra of the conventional antibracket.
Here $F$ is a general expression in $S$.
Let us consider the analogous transformation algebra in the general case.
The algebra corresponding to \mbox{$\sigma$} does not close in general,
but yields instead an algebra it is natural to call ``open'' (again a
terminology motivated by closed string field theory; see
ref. \cite{Zwiebach}, eqs.\ (4.60-4.61)):
\beq
[\delta_{\epsilon_{2}},\delta_{\epsilon_{1}}] F(S)=
\delta_{\epsilon_{3}} F(S)+
\widetilde{\Phi}_{T} \left(M(S) \otimes \epsilon_1 \otimes  \epsilon_2
\otimes e^{\otimes \frac{i}{\hbar}S} \right) F'(S) ~,
\label{open}
\eeq
with
\bea
\epsilon_3 &~\equiv~& \widetilde{\Phi}_{T} \left(\epsilon_1 \otimes
\epsilon_2 \otimes e^{\otimes \frac{i}{\hbar}S} \right) ~.
\label{e3open}
\eea

\noi
The additional terms on the right hand side of (\ref{open}) are here to
be understood as ``equation of motion'' terms,
and the gauge algebra is then of the usual open kind.
In the conventional case of
vanishing three-bracket, the ``equation of motion" term in (\ref{open})
drops out, and (\ref{e3open}) boils down to
\mbox{$\epsilon_3=\Phi_{T}^2 (\epsilon_1 \otimes  \epsilon_2) $}, thereby
reproducing (\ref{2closed}).

\noi
The easiest way to derive eq. (\ref{open}) is by using the main identity
(\ref{TSmainid}),
\bea
0 &=& \widetilde{\Phi}_{T}  \circ b_{\textstyle \widetilde{\Phi}_{T} }
\left(\epsilon_1 \otimes  \epsilon_2 \otimes
e^{\otimes \frac{i}{\hbar}S} \right) \cr
&=& \widetilde{\Phi}_{T} \left( \widetilde{\Phi}_{T} \left(
\epsilon_1 \otimes e^{\otimes \frac{i}{\hbar}S} \right)
\otimes  \epsilon_2 \otimes  e^{\otimes \frac{i}{\hbar}S} \right)
- \widetilde{\Phi}_{T} \left( \widetilde{\Phi}_{T} \left(
\epsilon_2\otimes e^{\otimes \frac{i}{\hbar}S} \right)
\otimes  \epsilon_1 \otimes  e^{\otimes \frac{i}{\hbar}S} \right)\cr
&&+ \widetilde{\Phi}_{T} \left( \widetilde{\Phi}_{T} \left(
\epsilon_1 \otimes  \epsilon_2 \otimes e^{\otimes \frac{i}{\hbar}S} \right)
\otimes  e^{\otimes \frac{i}{\hbar}S} \right)
+ \widetilde{\Phi}_{T} \left( \widetilde{\Phi}_{T} \left(
e^{\otimes \frac{i}{\hbar}S} \right)
\otimes \epsilon_1 \otimes  \epsilon_2 \otimes
e^{\otimes \frac{i}{\hbar}S} \right)
\label{fourtermmess}~,
\eea
and noting that
\beq
\frac{i}{\hbar} \delta_{\epsilon_{2}}\delta_{\epsilon_{1}}S=
 \frac{i}{\hbar}  \widetilde{\Phi}_{T} \left( \delta_{\epsilon_{2}}S
\otimes  \epsilon_1 \otimes  e^{\otimes \frac{i}{\hbar}S} \right)
= \widetilde{\Phi}_{T} \left( \widetilde{\Phi}_{T} \left(
\epsilon_2\otimes e^{\otimes \frac{i}{\hbar}S} \right)
\otimes  \epsilon_1 \otimes  e^{\otimes \frac{i}{\hbar}S} \right)~.
\eeq
Interestingly, the algebra can be made to close by
choosing the transformations \mbox{$\bar{\sigma}$} instead.
As we have emphasized before, the two transformations \mbox{$\bar{\sigma}$}
and  \mbox{$\sigma$} are equal ``on-shell":
\beq
   \bar{\sigma}(\epsilon) = \sigma(\epsilon) + M(S) \epsilon~.
\label{onshellequal}
\eeq
The closed algebra corresponding to  \mbox{$\bar{\sigma}$} is:
\beq
[\bar{\delta}_{\epsilon_{2}},\bar{\delta}_{\epsilon_{1}}] F(S)=
\bar{\delta}_{\epsilon_{3}} F(S)~,
\label{closed}
\eeq
with
\bea
\epsilon_3 &~\equiv~& e^{-\frac{i}{\hbar}S}\epsilon_1
\stackrel{\rightarrow}{T}
\epsilon_2 \ e^{\frac{i}{\hbar}S}
 - e^{-\frac{i}{\hbar}S}\epsilon_2  \stackrel{\rightarrow}{T}
\epsilon_1 \ e^{\frac{i}{\hbar}S}\cr
&~=~&{i \over \hbar} \epsilon_1\bar{\delta}_{\epsilon_{2}}S
-{i \over \hbar} \epsilon_2\bar{\delta}_{\epsilon_{1}}S
= \epsilon_1\bar{\sigma}(\epsilon_{2})
- \epsilon_2\bar{\sigma}(\epsilon_{1}) ~.
\eea
This holds even without assuming nilpotency of $T$.

\noi
Having found new nilpotent operators $\sigma$ and $\bar{\sigma}$ generated
by nilpotent $T$-operators, it is natural to consider the higher antibrackets
generated by $\sigma$ or $\bar{\sigma}$.
\beq
\Phi^n_{\bar{\sigma}} ( A_1 \otimes \ldots \otimes A_n)
=\widetilde{\Phi}_{T} \left(  e^{\otimes \frac{i}{\hbar}S}
\otimes A_1 \otimes \ldots \otimes A_n\right)~.
\label{bracbarsigma}
\eeq
This is the natural generalisation of the BRST operator to more entries.

\noi
Apart from the zero-bracket, the two sets of higher brackets
\mbox{$\Phi^n_{\sigma}$}, \mbox{$\Phi^n_{\bar{\sigma}}$} are equal:
\beq
\Phi^n_{\sigma}=\Phi^n_{\bar{\sigma}}~,~~~n \neq 0~,
\eeq
because the difference \mbox{$\sigma-\bar{\sigma}$} is a (left)
multiplication operator(cf.\ (\ref{onshellequal})).
Note that
\beq
\sigma ~=~ \Phi^1_{\bar{\sigma}} ~=~ \Phi^1_{\sigma}~.
\eeq

\noi
The careful reader will have noticed that each time \mbox{$\sigma$} was
treated in the past two sections,
we chose, whenever possible, arguments that did not involve the assumption
of a product for the algebra ${\cal A}$. For instance (\ref{fourtermmess})
could be derived easier with the help of (\ref{bracbarsigma}) and
(\ref{leiblaplanti}).

\noi
To summarize, the benefits of $\sigma$ are chiefly
that it can be written purely in terms of higher brackets,
$i.e.$  without the use of a product\footnote{A point crucial for
understanding why it so far has been $\sigma$ only which has surfaced
in closed string field theory. Interestingly, $\sigma$ is not seen as
a BRST transformation in closed string field theory, but rather as a
gauge transformation. We will return to this point in section 3.},
while $\bar{\sigma}$ have the nicest properties with respect to nilpotency,
invariance of master equation and closure of the transformation algebra.

\subsection{\sc  Finite Transformations}

\noi
If we keep the perspective that $\sigma$ and $\bar{\sigma}$ can be seen
as valid deformations $\delta S$ of a solution $S$ to the Master Equation
$M(S) = 0$, it is natural to ask for the analogous {\em finite}
deformations of $S$. In the case of the conventional Batalin-Vilkovisky
formalism, this has also been considered by Hata and Zwiebach in
ref. \cite{covariant}. We shall here consider the general case.
If we focus on $\sigma$, it is actually possible to derive, without too much
effort, an integrated version. By a curious twist of events, precisely
this case has been considered earlier in the context of closed string
field theory as well \cite{chrsch}. We shall here present a more direct
construction, making use of the machinery we derived in the previous
subsections. Consider infinitesimal transformation of the form
\beq
\delta_{\epsilon}A
= \sigma(\epsilon)
= \sum_{n=0}^{\infty} {1 \over {n!}}
\Phi^{n+1}_T( \epsilon  \otimes A^{\otimes n} )
= \widetilde{\Phi}_{T}(\epsilon  \otimes e^{\otimes A} )  ~.
\eeq
Here $\epsilon$ and $T$ are supposed to have the same Grassmann parity
and $A$ is bosonic.
The above transformations correspond, as mentioned previously, to gauge
transformations in closed string field theory \cite{Zwiebach} (there
with $A= \kappa \Psi$ being a string field).
The transformation parameter \mbox{$\epsilon \equiv \epsilon_0 \ dt$}
can be split into a
finite constant $\epsilon_0$ of same Grassmann parity as $\epsilon$,
and a bosonic infinitesimal parameter $dt$.
We want to integrate up this expression to finite transformations.
It follows that we have a 1.\  order initial value problem:
\bea
{d \over {dt}}  A(t) &=& \widetilde{\Phi}_T(\epsilon_0
\otimes e^{\otimes A(t)} ) \cr
&&\cr
A(t=0)&=&A_0~.
\eea
This can be rewritten as an integral equation
\beq
 A(t) = A_0 + \int_0^t ds   \widetilde{\Phi}_T(\epsilon_0
\otimes e^{\otimes A(s)} )~.
\eeq
Let us define \mbox{$a(t) \equiv e^{\otimes A(t)}$}.
Exponentiating the integral equation yields:
\beq
 a(t)= a_0 \otimes \exp \int_0^t ds
\widetilde{\Phi}_T(\epsilon_0  \otimes a(s) )~.
\eeq
Iterating this ``fixed-point integral equation" infinitely many times gives:
\bea
 a(t_1)&=& a_0 \otimes  \exp \int_0^{t _1} dt_2
\widetilde{\Phi}_T \left(\epsilon_0  \otimes
a_0 \otimes  \exp \int_0^{t _2} dt_3
 \widetilde{\Phi}_T \left(\epsilon_0  \otimes \ldots \right.\right. \cr
&& ~~~~~~~~~~~~~~~~~~
 \left.\left.\ldots \otimes a_0 \otimes \exp \int_0^{t _n} dt_{n+1}
\widetilde{\Phi}_T \left(\epsilon_0  \otimes \ldots
 \right) \ldots\right)\right)~.
\eea
Projecting $A(t_1)= \pi_{\cal A} \  a(t_1)$ to the original algebra
${\cal A}$ results in
\bea
 A(t_1)&=& A_0 +  \int_0^{t _1} dt_2
\widetilde{\Phi}_T \left(\epsilon_0  \otimes
a_0 \otimes \exp \int_0^{t _2} dt_3
\widetilde{\Phi}_T \left(\epsilon_0   \otimes \ldots  \right.\right. \cr
&& ~~~~~~~~~~~~~~~~~~
 \left.\left.\ldots \otimes a_0 \otimes \exp \int_0^{t _n} dt_{n+1}
\widetilde{\Phi}_T \left(\epsilon_0  \otimes \ldots
 \right) \ldots\right)\right)~.
\label{infinteq}
\eea
Note that one only has to apply the fixed-point integral equation $n$ times,
to get the $n$'th order contribution with respect to the transformation
parameter $\epsilon_0$. 
The first few orders in the parameter $\epsilon_0$ are:
\bea
A(\epsilon_0) &=&A_0 +  \widetilde{\Phi}_T(\epsilon_0  \otimes a_0 )
+{1 \over 2}  \widetilde{\Phi}_T \left( \epsilon_0  \otimes
a_0 \otimes  \widetilde{\Phi}_T(\epsilon_0  \otimes a_0) \right) \cr
&& \cr &&+{1 \over 6}  \widetilde{\Phi}_T \left( \epsilon_0  \otimes
a_0 \otimes  \widetilde{\Phi}_T(\epsilon_0  \otimes a_0)
\otimes  \widetilde{\Phi}_T(\epsilon_0  \otimes a_0)  \right)\cr
&& \cr &&+{1 \over 6}  \widetilde{\Phi}_T \left( \epsilon_0  \otimes
a_0 \otimes  \widetilde{\Phi}_T \left( \epsilon_0  \otimes
a_0 \otimes  \widetilde{\Phi}_T(\epsilon_0  \otimes a_0)  \right) \right) \cr
&& \cr &&+{1 \over 24}  \widetilde{\Phi}_T \left( \epsilon_0  \otimes
a_0 \otimes  \widetilde{\Phi}_T(\epsilon_0\otimes a_0)^{\otimes 3} \right) \cr
&& \cr &&+{1 \over 8}  \widetilde{\Phi}_T \left( \epsilon_0  \otimes
a_0 \otimes  \widetilde{\Phi}_T(\epsilon_0  \otimes a_0) \otimes
 \widetilde{\Phi}_T \left( \epsilon_0  \otimes
a_0 \otimes  \widetilde{\Phi}_T(\epsilon_0  \otimes a_0) \right)  \right) \cr
&& \cr &&+{1 \over 24}  \widetilde{\Phi}_T \left( \epsilon_0  \otimes
a_0 \otimes   \widetilde{\Phi}_T \left( \epsilon_0  \otimes
a_0 \otimes  \widetilde{\Phi}_T(\epsilon_0  \otimes a_0)^{\otimes 2}
\right) \right)\cr
&& \cr &&+{1 \over 24}  \widetilde{\Phi}_T \left( \epsilon_0  \otimes
a_0 \otimes   \widetilde{\Phi}_T \left( \epsilon_0  \otimes
a_0 \otimes   \widetilde{\Phi}_T \left( \epsilon_0  \otimes
a_0 \otimes  \widetilde{\Phi}_T(\epsilon_0  \otimes a_0) \right)  \right)
\right)
+{\cal O}\left((\epsilon_0)^5 \right)~.
\eea

\noi
Although eq. (\ref{infinteq}) gives the finite transformation in closed
form by taking the limit $n \to \infty$, it is clearly not very useful
beyond the expansion in $\epsilon_0$ (illustrated to ${\cal O}(\epsilon_0^5)$
above). It is therefore of more interest to consider the order-by-order
expansion. Let us first comment on the type of terms that can arise.
Besides the zeroth-order term $A_0$,
all terms begin (and {\em end}) with a bracket \mbox{$\widetilde{\Phi}_T$},
i.e.\ two brackets are never multiplied at the lowest level of nesting.
Note that the symmetry factor ${1 \over 8}$ in the above expression
breaks the otherwise apparent factorial pattern of the first orders, so
the rule for giving the coefficients is clearly not that simple.
In general, the symmetry factor for a term
can be deduced according to two simple rules found empirically in
ref. \cite{chrsch}, and which easily can be read off from formula
(\ref{infinteq}). The rule is the following. For each bracket
\mbox{$\widetilde{\Phi}_T$} appearing in the considered term, do the
following:
\begin{itemize}
\item
If $k$ entries are equal, divide by ${1 \over {k!}}$.
\item
Divide by the total number $N$ of $\epsilon_0$'s 
appearing {\em somewhere} inside the bracket 
($i.e.$ also the $\epsilon_0$'s in further nested brackets).
\end{itemize}
These two simple rules suffice in determining the whole expansion. Of course,
one can as easily simply expand eq. (\ref{infinteq}).

\setcounter{equation}{0}
\section{Connection to Closed String Field Theory}

\noi
Non-polynomial closed string field theory \cite{Kugo,Zwiebach} is based on
a so-called ``string product'' which shares a number of properties with
higher antibrackets. This is particularly obvious in the conventions of
Zwiebach \cite{Zwiebach}, which we will follow here. For an arbitrary
genus $g$, the $n$th string product is denoted by $[A_1,\ldots,A_n]_g$.
It has $n$ entries of states (string fields) $A_i$, and it maps these states
into a new state.\footnote{In closed string field theory, 
these states are assumed to be annihilated by certain operators
$b_0^-$ and $L_0^-$ (a property the string product inherits), but this
assumption is not required in the following considerations, when restricted
to properties of the string products alone.} This string
product is supercommutative,
\beq
[A_1,\ldots,A_{i-1},A_i,\ldots,A_n]_g = (-1)^{\epsilon_{A_{i-1}}
\epsilon_{A_{i}}}[A_1,\ldots,A_i,A_{i-1},\ldots,A_n]_g ~,
\eeq
and Grassmann-odd:
\beq
\epsilon([A_1,\ldots,A_n]_g) = \sum_{i=1}^n\epsilon_{A_{i}} + 1 ~.
\eeq
The string product also carries ghost number (the same for any genus $g$),
but this notion is not of importance for what follows. In addition
to the string product, an important r\^{o}le is played by the BRST
operator $Q$.

\noi
In classical closed string field theory, corresponding to genus zero, the
string product satisfies a so-called ``main identity'' of the form
\cite{Sen,Zwiebach}
\bea
0 &=& Q[A_1,\ldots,A_n]_0 + \sum_{i=1}^n(-1)^{\epsilon_{A_{1}}+\cdots
\epsilon_{A_{i-1}}}[A_1,\ldots,QA_i,\ldots,A_n]_0 \cr
&&+ \sum_{\{i_l,j_k\}}\sigma(i_l,j_k)\left[A_{i_{1}},\ldots,A_{i_{l}},
[A_{j_{1}},\ldots,A_{j_{k}}]_0\right]_0 ~,\label{mainid}
\eea
where the last sum is restricted to $l\geq 1, k\geq 2$, and $l+k = n$.
The sign factor $\sigma(i_l,j_k)$ is what is picked up by the
prescribed reordering of terms, using the fact the string product is
supercommutative.

\noi
The BRST operator $Q$ is defined on a given conformal background, and the
whole string field theory is then also defined on such a background. At
genus zero, this means that the ``zero-product'' corresponding to $n=0$
must be taken to vanish:
\beq
[~\cdot~]_0 ~=~ 0 ~.\label{zeroprod}
\eeq
(The corresponding definition away from a conformal background will be
discussed later.) The first non-trivial string product is thus the
``one-product'', a linear map that takes one string state into another.
It is given by
\beq
[A]_0 ~\equiv~ QA ~.
\eeq

\noi
The classical non-polynomial closed string field theory action can then be
written \cite{Kugo,Zwiebach}
\beq
S(\Psi) ~=~ \frac{1}{\kappa^2}\sum_{n=2}^{\infty}\frac{\kappa^n}{n!}
\{\Psi,\ldots,\Psi\}_0 ~, \label{SFTaction}
\eeq
where $\{A,B_1,\ldots,B_n\}_0$, a Grassmann-even $(n+1)$-bracket,
is defined by an inner product,
\beq
\{A,B_1,\ldots,B_n\}_0 ~\equiv~ \langle A,[B_1,\ldots,B_n]_0\rangle ~,
\eeq
with the following exchange relation:
\beq
\langle A,B \rangle = (-1)^{(\epsilon_A+1)(\epsilon_B+1)} \langle B,A
\rangle ~.
\eeq
A more familiar expression for the closed string field theory action is
obtained by using $\{\Psi,\Psi\}_0 \equiv \langle\Psi,[\Psi]_0\rangle
= \langle\Psi,Q\Psi\rangle$ \cite{Kugo,Zwiebach}:
\beq
S(\Psi) = \frac{1}{2}\langle\Psi,Q\Psi\rangle + \frac{1}{\kappa^2}
\sum_{n=3}^{\infty}\frac{\kappa^n}{n!}\{\Psi,\ldots,\Psi\}_0 ~,
\eeq
where the last bracket has $n$ entries. The classical equations of motion
then take the form
\beq
Q\Psi + \frac{1}{\kappa}\sum_{n=2}^{\infty}\frac{\kappa^n}{n!}
[\Psi,\ldots,\Psi]_0 
~=~ \frac{1}{\kappa}\sum_{n=1}^{\infty}\frac{\kappa^n}{n!}
[\Psi,\ldots,\Psi]_0 ~=~ 0 ~.\label{stringeqm.}
\eeq
Finally, the closed string field theory action (\ref{SFTaction}) is left
invariant by the following gauge transformations:
\beq
\delta_{\epsilon}\Psi 
~=~ Q\epsilon + \sum_{n=1}^{\infty}\frac{\kappa^n}{n!}
[\Psi,\ldots,\Psi,\epsilon]_0 ~=~ \sum_{n=0}^{\infty}\frac{\kappa^n}{n!}
[\Psi,\ldots,\Psi,\epsilon]_0 ~.\label{sftgauge}
\eeq

\noi
If we compare these string field theory expressions with the identities
among higher antibrackets
we derived in the previous sections, it is tempting to identify the $n$th
string product at genus zero with the $n$th
antibracket generated by an odd operator  $T$:
\beq
[A_1,\ldots,A_n]_0 ~=~ \Phi^n_T(A_1,\ldots,A_n) ~. \label{stringbracket}
\eeq
The obvious obstruction to such an identification is the lack of a simple
product ``$~\cdot~$'' of, using the notation of section 2, the algebra
${\cal A}$. Still, let us consider the similarities. 
We have already listed the pertinent
properties of the string products. All of these properties are shared
with the higher antibrackets: They are both Grassmann-odd, graded
commutative under exchange of entries, and the crucial "main identity"
of the string products is recognized as being identical to the
identity (\ref{mid}) of higher antibrackets. 

\noi

Consider now the equations of motion (\ref{stringeqm.}), which
in the previous section played the r\^{o}le of the full Master Equation
(the action $S$ replacing the string field $\Psi$). We can view this
equation, with its infinite sum of higher brackets (or string products),
from two points of view. Either as a clever way of representing
the particular combination of exponential functions without reference to
the algebra by means of which these exponential functions could be
defined, or as a very complicated way of writing the simple formula
\beq
e^{-\kappa\Psi}\stackrel{\rightarrow}{T}e^{\kappa\Psi} ~=~ 0
\eeq
through its power series expansion. Of course, to give meaning to
$\exp(\kappa\Psi)$, we would have to assume that it is possible to redefine 
ghost
number assignments so that $\kappa\Psi$ becomes of ghost number zero.
Closed string field theory is tied to the formulation in terms
of a power series expansion.

\noi
Similarly, the gauge symmetry of string field theory (\ref{sftgauge})
can be understood as the infinite-series expansion of the simple expression
\beq
\delta_{\epsilon}\Psi ~=~ e^{-\kappa\Psi}\left[\stackrel{\rightarrow}{T},
\epsilon \right]e^{\kappa\Psi} ~.
\eeq
We have also already seen the usefulness of the higher-antibracket
formalism when deriving what in closed string field theory is viewed
as the analogous finite gauge transformations (in section 2.9). In all
of these cases, we can use the algebra ${\cal A}$ to derive results
with far greater ease, and whenever these results are expressible in
terms of higher antibrackets alone (without using the new product) we
find that the expressions coincide with those of closed string field
theory. 

\noi
It is also of interest to see the results of subsection 2.8 from the
point of view of closed string field theory. It was noted by
Ghoshal and Sen \cite{Ghoshal} that there is an apparent clash between
the gauge symmetry of closed string field theory being {\em open} off-shell,
while the gauge transformations of the low-energy effective field theory
derived from this theory form an algebra which {\em closes} off-shell.
By analyzing special cases, they found that the usual gauge
transformations of closed string field theory combine with ``trivial''
gauge transformations (proportional to the equations of motion) to give
the proper transformations (which close) of the low-energy theory.
Ghoshal and Sen in fact
conjecture that all gauge transformations of closed string field
theory can be organized in such a manner (by adding suitable ``equation
of motion terms'') that the algebra eventually closes off-shell.
Our symmetry operator $\bar{\sigma}$ is precisely of this
kind, but it cannot as it stands be given an interpretation in closed
string field theory, since it -- in contrast to $\sigma$ -- involves
the product of string fields discussed above.

\subsection{\sc Beyond Conformal Backgrounds}

\noi
An interesting place for considering the analogy between string products
and higher antibrackets
is that of closed string field theory in a background
that is not conformal. Zwiebach \cite{Zwiebach} has analyzed the fate
of the string product algebra in this situation.

\noi
So far the analogy has been based on the assumption
that the ``zero-product'' (\ref{zeroprod}) is vanishing. Away from
a conformal background this zero-product will no longer vanish. Zwiebach
calls it $F$, and distinguishes the new string products by a prime
\cite{Zwiebach}:
\beq
[~\cdot~]'_0 ~=~ F ~.\label{nczeroprod}
\eeq
Denoting, accordingly, also the new BRST-like operator by $Q'$, some of the
first few identities that generalize the ``main identity'' of eq.
(\ref{mainid}) read \cite{Zwiebach}:
\bea
Q'F &~=~& 0 \\ \label{QprimeF}
Q'^2A + [F,A]'_0 &=& 0 \label{Qprime2}
\eea
This last equation (\ref{Qprime2}) gives the violation of $Q'$-nilpotency
away from a conformal background. The analogue of $Q$ differentiating
the two-product (\ref{dfg}) becomes
\beq
Q'[A_1,A_2]'_0 + [Q'A_1,A_2]'_0 + (-1)^{\epsilon_{A_{1}}}[A_1,Q'A_2]'_0
+ [F,A_1,A_2]'_0 = 0 ~,
\eeq
and the higher-order identities can also be worked out.

\noi
The appearance of a non-trivial zero-product $F$ has a completely
natural explanation in terms of higher antibrackets: it corresponds to
the inclusion of a non-trivial zero-bracket $\Phi^0_T$.

\noi
In closed string field theory, one studies the behavior away from a
conformal background by shifting the string field:
\beq
\Psi ~\to~ \Psi_0 + \Psi ~,
\eeq
where $\Psi_0$ does {\em not} solve the classical equation of
motion.\footnote{The analogous study of shifts with a $\Psi_0$ that still
solves the equations of motion (corresponding to a new, but still conformal,
background), was first considered by Sen \cite{Sen}. In this case the
zero-product must still be taken to vanish.} The precise connection
between such a shift and the emergence of a new algebra of string brackets
that now involves the zero-product is easily understood if one accepts
the formulation in terms of higher antibrackets and the new, assumed,
string product. Consider the equation of motion for the unshifted field.
We can write it as
\bea
0 ~=~ e^{-\kappa\Psi}\stackrel{\rightarrow}{T}e^{\kappa\Psi}
&~=~& e^{-\kappa(\Psi - \Psi_0) -
\kappa\Psi_0}\stackrel{\rightarrow}{T}e^{\kappa\Psi_0
+ \kappa(\Psi - \Psi_0)} \cr
&~=~& e^{-\kappa(\Psi - \Psi_0)}\stackrel{\rightarrow}{T}{}'
e^{\kappa(\Psi - \Psi_0)} ~,
\eea
that is, the equation of motion for the shifted string field $\Psi -
\Psi_0$, {\em with respect to a new  nilpotent BRST operator},
\beq
\stackrel{\rightarrow}{T}{}' ~\equiv~
e^{-\kappa\Psi_0}\stackrel{\rightarrow}{T} e^{\kappa\Psi_0} (\cdot)~,
\eeq
a conjugate version of  \mbox{$\stackrel{\rightarrow}{T}$}.
Recall that \mbox{$\stackrel{\rightarrow}{T}$} here acts on everything to
its right.

\noi
Because $\Psi_0$ is assumed not to solve the classical equation of motion,
it follows immediately that the new string product algebra will have a
non-trivial zero-product:
\bea
\stackrel{\rightarrow}{T}{}'  1   &~=~&
e^{-\kappa\Psi_0}\stackrel{\rightarrow}{T}
 e^{\kappa\Psi_0}  \cr
&~=~& \sum_{n=1}^{\infty}\frac{\kappa^n}{n!}
[\Psi_0,\ldots,\Psi_0]_0 ~\neq ~ 0 ~,
\eea
$i.e.$, precisely ($\kappa$ times) the left hand side of the equation of
motion for $\Psi_0$ (which by assumption is non-vanishing). We identify
it as $F = [~\cdot~]_{0}'$ above.  Note, incidentally, that
\beq
\stackrel{\rightarrow}{T}{}'F = \stackrel{\rightarrow}{T}{}'
\stackrel{\rightarrow}{T}{}'(1) = 0 ~,
\eeq
but this identity is {\em not} the same as eq. (\ref{QprimeF}). 
The ``BRST-like" operator $Q'$ is the one-bracket $\Phi^1_{T'}$ 
associated with $\stackrel{\rightarrow}{T}{}'$, not
the BRST operator $\stackrel{\rightarrow}{T}{}'$ itself.

\noi
The whole sequence of main identities can of course now be rewritten in
terms of $\stackrel{\rightarrow}{T}{}'$, rather than $\stackrel{\rightarrow}
{T}$. The only new feature compared with the usual main identities of
closed string field theory is that the 0-bracket $\Phi^0$ is non-vanishing.
In particular, one sees immediately that the first identities (\ref{QprimeF})
and (\ref{Qprime2}) are trivially included in eq. (\ref{mid}), and similarly
for the higher identities. They are all contained in eq. (\ref{mid}).

\noi
Note that $\stackrel{\rightarrow}{T'}$ is nilpotent simply as
a consequence of $\stackrel{\rightarrow}{T}$ being nilpotent. It can be
viewed as a genuine BRST operator corresponding to the shifted background.
The ``BRST-like''
operator $Q'$ of closed string field theory \cite{Zwiebach} is in the
present context rather seen as the one-bracket; it is not nilpotent when
$F \neq 0$.

\noi
We have thus shown that when shifting the string field $\Psi$  by $\Psi_0$,
almost all of the formalism remains intact, and in particular almost
everything
can eventually be expressed in terms of string products. It is therefore
not surprising that
these results can also be derived directly on the basis of string products
alone \cite{Zwiebach}. They just appear with far more ease in the present
picture. There are also interesting exceptions, such as the new nilpotent
BRST operator $\stackrel{\rightarrow}{T'}$. This is the appropriate BRST
operator for shifted backgrounds, but it cannot be expressed solely
in terms of antibrackets (or string products), and therefore has no
obvious analogue in closed string field theory.

\setcounter{equation}{0}
\section{An $Sp(2)$-Symmetric Formulation}

\noi
As discussed in section 2, the higher antibrackets give rise to a BRST
symmetry which is a generalization of the BRST symmetry of Batalin and
Vilkovisky. An obvious question to ask is whether one analogously can 
find a formulation that includes both BRST symmetry and anti-BRST symmetry. 
There have been various suggestions for Lagrangian BRST formulations 
\`{a} la Batalin and Vilkovisky which includes 
the extended BRST--anti-BRST symmetry. 
All these have from the outset included the BRST--anti-BRST symmetries in
an $Sp(2)$ symmetry. The original approach is due to Batalin, Lavrov
and Tyutin \cite{BLT}, and it has recently been suggested that this
formulation be rephrased in terms of what has been called ``triplectic
quantization'' \cite{BM}\footnote{For alternative viewpoints, suggestions
for alternative but physically equivalent schemes, and a
derivation of the relation between the two different schemes, see ref.
\cite{sp2}.}

\noi
The main new ingredient of an $Sp(2)$-symmetric formulation of
conventional Lagrangian quantization is a Grasmann-odd
vector field $V$, which satisfies
$V^2 = 0$, and which must be added to the $\Delta$-operator. We take
$V$ to be a differential operator based on a right-derivative. In the
original formulation of ref. \cite{BLT}, the following relations are
assumed:
\begin{eqnarray}
V^{\{a}V^{b\}} ~&=&~ 0 \label{VV}\\
\Delta^{\{a}\Delta^{b\}} ~&=&~ 0 \label{DeltaDelta}\\
\Delta^{\{a}V^{b\}} + V^{\{a}\Delta^{b\}} ~&=&~ 0 ~.\label{DeltaV}
\end{eqnarray}
Here $a, b, \ldots$ denote indices in $Sp(2)$, the invariant tensor of
which is $\epsilon_{ab}$.\footnote{In the first formulation of the
$Sp(2)$-symmetric triplectic quantization \cite{BM}, the condition
(\ref{DeltaV}) was required to hold even before antisymmetrization in
the $Sp(2)$-indices. However, as was noted in ref. \cite{ND}, the
above more general condition is all that is required. See also
ref. \cite{BM1}.} Symmetrization in $Sp(2)$ indices is defined by
\beq
F^{\{a} G^{b\}} ~\equiv~ F^a G^b + F^b G^a ~,
\eeq
and these indices are raised and lowered by the $\epsilon$-tensor.

\noi
In refs. \cite{BM,ND,BM1} the $\Delta^a$-operators
are assumed to of purely 2nd order, while the $V^a$-operators are
assumed to be of purely 1st order. However, in actual applications it
is usually the combinations
\beq
\Delta^a_{\pm} ~\equiv~ \Delta^a \pm \frac{i}{\hbar}V^a \label{Deltapm}
\eeq
which appear. This suggests that we should simply view $\Delta^a_{\pm}$
as more general 2nd-order odd differential operators (still excluding
a constant term). It follows from (\ref{VV}), (\ref{DeltaDelta}) and
(\ref{DeltaV})
that
\beq
\Delta^{\{a}_{\pm} \Delta^{b\}}_{\pm} ~=~ 0 ~.\label{Sp2nilp}
\eeq
Since by definition $V^a$ is of first order, the
antibrackets defined by use of either $\Delta^a$ or $\Delta^a_{\pm}$
will coincide. These antibrackets are born with an $Sp(2)$-index:
\beq
(F,G)^a \equiv (-1)^{\epsilon_G}\Delta^a(FG) - \Delta^a(F)G -
(-1)^{\epsilon_G}F\Delta^a(G) ~.
\eeq

\noi
The above antibrackets satisfy the usual exchange relation (\ref{exchange}),
and the same graded Leibniz rules (\ref{Leibniz}). The analogue of
the graded Jacobi identity (\ref{Jacobi}) reads
\beq
\sum_{\mbox{\rm cycl.}}(-1)^{(\epsilon_F+1)(\epsilon_H+1)}
(F,(G,H)^{\{a})^{b\}} ~=~ 0 ~,\label{Jacobisp2}
\eeq
and the $Sp(2)$-covariant version of the relation (\ref{dfg}) is
\beq
\Delta^{\{a}(F,G)^{b\}} ~=~ (F,\Delta^{\{a}G)^{b\}} - (-1)^{\epsilon_G}
(\Delta^{\{a}F,G)^{b\}} ~.\label{dfgsp2}
\eeq
Furthermore, it follows from the above definitions that the vector
fields $V^a$ differentiate the antibrackets according to
\beq
V^{\{a}(F,G)^{b\}} ~=~ (F,V^{\{a}G)^{b\}} -
(-1)^{\epsilon_G}(V^{\{a}F,G)^{b\}} ~.
\eeq
This implies that also the relation (\ref{dfgsp2}) remains valid if we
replace $\Delta^a$ by $\Delta^a_{\pm}$.

\noi
Our task is now to generalize the above construction to the case of
higher antibrackets. The obvious starting point is to introduce two
higher-order $\Delta^a$-operators, and proceed as in section 2, using
the $Sp(2)$-algebra (\ref{DeltaDelta}). The analogous $V^a$-operator,
taken by definition to be always of 1st order, can be introduced trivially
by letting $\Delta^a \to \Delta^a_{\pm} \equiv \Delta^a \pm (i/\hbar)
V^a$, where $V^a$ simply equals the 1st-order part of $\Delta^a$.
The main ingredient is therefore the existence of two odd
differential operators of arbitrary order, and with the algebra
of $\Delta^a$ as in (\ref{DeltaDelta}). As in the previous section, we
can include the case of a possibly non-vanishing constant pieces in
these differential operators as well, corresponding to $\Delta^a(1)$
not necessarily being zero.

\noi
\subsection{\sc Sp(2)-Covariant Higher Antibrackets}

\noi
All necessary ingredients for the extension of the above
$Sp(2)$-symmetric formulation to the higher-antibracket BRST symmetry
have been given in section 2. In particular, we refer to section
2.4, where we gave the algebra of higher antibrackets generated by
{\em two} nilpotent operators $S, T$. In accordance with the above
formalism, we shall here denote these operators by $\Delta^a$ and
$\Delta^b$. All the subsequent manipulations remain valid if we
replace these by $\Delta_{\pm}$ through the definition (\ref{Deltapm}).
Because of the proliferation of indices, we drop the subscript
$_{\Delta}$ on the higher antibrackets, and just indicate the relevant
$\Delta$-operator by its $Sp(2)$-index $a$. For simplicity, we
take
\beq
\Delta^a(1) ~=~ 0 ~,
\eeq
so that there are no zero-brackets 
(they can of course trivially be included).

\noi
With the operators $\Delta^a$ being Grassmann-odd, the algebra of
$Sp(2)$-symmetric higher antibrackets can then be written:
\beq
\sum_{r=1}^n {1\over {r!(n-r)!}}\sum_{\pi \in {\cal S}_n }
(-1)^{\epsilon_{\pi}} \Phi^{n-r+1}_{\{a}\left(
\Phi^r_{b\}}  (A_{\pi(1)} \otimes \ldots \otimes A_{\pi(r)}  )
 \otimes A_{\pi(r+1)} \otimes \ldots \otimes A_{\pi(n)} \right) = 0 ~.
\eeq
This algebra contains all the usual identities of $Sp(2)$-symmetric
quantization as outlined above, and the appropriate generalization if
higher antibrackets are included. In detail, the first identity is
nothing but $Sp(2)$-nilpotency of the operators $\Delta^a$ (cf.
eq. (\ref{Sp2nilp}):
\beq
\Phi^1_{\{a}(\Phi^1_{b\}}(A)) ~=~ 0 ~,
\eeq
while the 2nd identity gives the $Sp(2)$-covariant rule for how
$\Delta^a$ differentiates the ``two-antibracket'' (as in eq.
(\ref{dfgsp2})):
\beq
\Phi^1_{\{a}(\Phi^2_{b\}}(A_{1} \otimes A_{2}))
= -\Phi^2_{\{a}\left(\Phi^1_{b\}}(A_{1})\otimes A_{2}\right) -
(-1)^{\epsilon_{A_{1}}\epsilon_{A_{2}}}\Phi^2_{\{a}\left(
\Phi^1_{b\}}(A_{2})\otimes A_{1}\right) ~.
\eeq
(Note that this identity is not altered by the presence of higher order
operators in the $\Delta^a$'s).
The next identity is the $Sp(2)$-covariant analogue of the Jacobi
identity (\ref{Jacobisp2}), including its possible breaking when the
$\Delta^a$'s are of order 3 or higher:
\begin{eqnarray}
&&\Phi^2_{\{a}\left(
\Phi^2_{b\}} \left(A_1 \otimes A_2 \right)
 \otimes  A_3 \right)
+(-1)^{\epsilon_{A_1}(\epsilon_{A_2}+\epsilon_{A_3})}
\Phi^2_{\{a}\left(
\Phi^2_{b\}} \left(A_2 \otimes A_3 \right)
 \otimes  A_1 \right) \cr
&&+(-1)^{(\epsilon_{A_1}+\epsilon_{A_2})\epsilon_{A_3}}
\Phi^2_{\{a}\left(
\Phi^2_{b\}} \left(A_3 \otimes A_1\right)
 \otimes  A_2 \right)
{}~=~ - \Phi^3_{\{a}\left(
\Phi^1_{b\}} \left(A_1 \right)
 \otimes A_2  \otimes A_3 \right)\cr
&&- (-1)^{\epsilon_{A_1}(\epsilon_{A_2}+\epsilon_{A_3})}
\Phi^3_{\{a}\left(
\Phi^1_{b\}} \left(A_2 \right)
 \otimes A_3  \otimes A_1 \right)
- (-1)^{(\epsilon_{A_1}+\epsilon_{A_2})\epsilon_{A_3}}
\Phi^3_{\{a}\left(
\Phi^1_{b\}} \left(A_3 \right)
 \otimes A_1  \otimes A_2 \right)\cr
&&- \Phi^1_{\{a}\left(
\Phi^3_{b\}} \left(A_1 \otimes A_2 \otimes A_3 \right) \right)   ~.
\end{eqnarray}
The subsequent identities are of course completely
new, involving higher and higher order of antibrackets. They can be read off
directly from eq. (\ref{mid}).
Also the higher main identity (\ref{newmain}) generates a series of new
identities.
We quote the first few:
\bea
0 &=&\Phi^{1}_{a}\left(  \Phi_{b}^{1} \left(
\Phi^{1}_{c} (A)  \right)  \right) ~.
\eea
This identity is rather trivial, but  it turns out to be very convenient for
proving that no other independent $Sp(2)$ main identities exist. The
next reads 
\bea
0 &=&\Phi^{1}_{a}\left(  \Phi_{b}^{1} \left(
\Phi^{2}_{c} (A_{1} \otimes A_{2})  \right)  \right)
+\Phi^{1}_{a}\left(  \Phi_{b}^{2} \left(
\Phi^{1}_{c} (A_{1} ) \otimes A_{2} \right)  \right) \cr && \cr
&&+(-1)^{\epsilon_{A_{1}}\epsilon_{A_{2}}}
\Phi^{1}_{a}\left(  \Phi_{b}^{2} \left(
\Phi^{1}_{c} (A_{2} ) \otimes A_{1} \right)  \right)
+\Phi^{2}_{a}\left(  \Phi_{b}^{1} \left(
\Phi^{1}_{c} (A_{1} )  \right) \otimes A_{2} \right) \cr && \cr
&&+(-1)^{\epsilon_{A_{1}}\epsilon_{A_{2}}}
\Phi^{2}_{a}\left(  \Phi_{b}^{1} \left(
\Phi^{1}_{c} (A_{2} )  \right) \otimes A_{1} \right)
+(-1)^{\epsilon_{A_{1}}}
 \Phi^{2}_{a}\left(  \Phi_{b}^{1} (A_{1})
\otimes \Phi^{1}_{c} (  A_{2})    \right)\cr && \cr
&&+(-1)^{(\epsilon_{A_{1}}+1)\epsilon_{A_{2}}}
 \Phi^{2}_{a}\left(  \Phi_{b}^{1} (A_{2})
\otimes \Phi^{1}_{c} (  A_{1})    \right) ~.
\eea
The interesting point about these new identities (the first few of which
of course are valid also in conventional $Sp(2)$ BRST quantization) is
that they do not involve symmetrizations in the $Sp(2)$ indices. The
higher identities can be read off from eq. (\ref{newmain}) in Appendix B.

These higher main identities are  qualified guesses for  what will
arise in
an $Sp(2)$ symmetric formulation of genus zero closed string field theory.

\noi
We next turn to the question of the corresponding BRST operators. In the
conventional $Sp(2)$-covariant scheme of ref. \cite{BLT}, one can show
\cite{sp2} -- as expected -- that the two symmetries are generated by the
two antibrackets and the solution to the Master Equations
\beq
\Delta^a_{+} \exp\left[\frac{i}{\hbar}S\right] ~=~ 0 ~.
\eeq
More interestingly, also in this context one derives a ``quantum BRST
operator'' (see the 2nd reference of \cite{sp2}), which reads
\beq
\sigma^a\epsilon ~=~ (\epsilon,S)^a + V^a\epsilon - (i\hbar)
\Delta^a\epsilon ~,
\eeq
where the first-order contribution $V^a$ to $\Delta^a$ explicitly
separates out.

\noi
Consider now the corresponding BRST operators in the generalized
situation in which one has higher antibrackets. Repeating the exercise
of the analogous situation without $Sp(2)$ symmetry, letting
\beq
T^{a}(F) = (-1)^{\epsilon_F} \Delta_{+}^{a}(F)~.
\eeq
one finds immediately that the appropriate generalization is (rescaling by a
factor
of $(i/\hbar)$, converting to left-derivatives, and lowering the $Sp(2)$
index):
\bea
\sigma_a\epsilon
&~=~& \widetilde{\Phi}_a(  \epsilon \otimes e^{\otimes {i \over \hbar} S})\cr
&~=~& \sum_{n=0}^{\infty} {1\over{n!}} \left({i \over \hbar} \right)^{n}
\Phi^{n+1}_{a} \left(  \epsilon \otimes S^{\otimes n} \right) \cr
&~=~& e^{-\frac{i}{\hbar}S}[\stackrel{\rightarrow}{T_a},
\epsilon]e^{\frac{i}{\hbar}S}~,
\label{sp2sigma}
\eea
and similarly for the associated BRST operator $\bar{\sigma}^a$, which one
can define
completely analogous to the case without $Sp(2)$ symmetry. When expanded
as a possibly infinite sum, the first
three terms of eq. (\ref{sp2sigma}) agree with the corresponding $Sp(2)$
quantum BRST operator of ref. \cite{sp2}. The new terms involve higher
and higher antibrackets precisely as anticipated. By construction,
\beq
\sigma_{\{a}\sigma_{b\}} ~=~ 0 ~,
\eeq
to all orders.

\setcounter{equation}{0}
\section{Conclusions}

\noi
Higher antibrackets provide us with a rich mathematical background
for studying various quantization problems in physics. They give the
obvious generalization of the Batalin-Vilkovisky formalism to situations
in which the $\Delta$-operator is of order 3 or higher. When viewed from
this more general perspective even the original Batalin-Vilkovisky
formalism is seen in a completely new light. Many of the ingredients
of the Lagrangian BRST formalism suddenly become very natural. For example,
in the conventional Batalin-Vilkovisky formalism the quantum Master Equation
involves both the conventional antibracket (the two-antibracket from
the present perspective) and the $\Delta$-operator. Usually, the need for
the quantum correction in the form of this $\Delta$-operator is viewed
as a kind of coincidence, the result of a particular correction from
the path integral measure to the classical BRST transformation of the
action. Similarly, the ``quantum correction'' to the classical BRST
transformation due to this $\Delta$-operator is seen as a (slightly
annoying) modification of the otherwise fully ``anticanonical'' formalism
that only involves the use of a two-antibracket: a Grassmann-odd analogue
of the Poisson bracket.
What we have seen here, is that the $\Delta$-operator is in no
way mysteriously present in the formalism. It plays two r\^{o}les: First,
it is the operator by which higher antibrackets are formed, and second,
it really is to be viewed as a ``one-antibracket'', completely on par with
the conventional antibracket. If $\Delta(1)$ would not vanish, this
identification would no longer hold. The quantum Master Equation is based
on $\Delta$, and it holds in all generality that this equation can be
expressed solely in terms of the higher antibrackets generated by $\Delta$.

\noi
The fact that an almost-canonical formulation\footnote{With respect to
an odd Poisson-like bracket, the usual antibracket.}
of the Lagrangian quantization
program exists, is thus in many respects coincidental, and not fundamental.
It is due to the fact that in the conventional representation of fields
and antifields the BRST operator of Schwinger-Dyson BRST symmetry (and
hence $\Delta$) is of 2nd order. In general, a Master Equation of the
form
\beq
\Delta \exp\left(\frac{i}{\hbar}S\right) = 0
\eeq
will contain an infinite series of arbitrarily high antibrackets. The
canonical considerations are of course limited to the two-antibracket.

\noi
{}From the Lagrangian BRST quantization point of view it is interesting that
the appearance of the $\Delta$-operator can be traced to a totally
different origin: that of integrating out ghosts while keeping the
antighosts \cite{AD}. Also from this point of view the $\Delta$-operator
immediately appears on an equal footing with the two-antibracket: the same
ghost integration that introduces the conventional antibracket in the
BRST operator also
simultaneously introduces the $\Delta$-operator. It is nevertheless
astonishing that the whole mathematical framework of higher antibrackets
can be derived by simple ghost-field integrations in the Lagrangian path
integral \cite{AD2,AD1}. The fact that there is an analogous construction
from the ghost momentum representation of Hamiltronian BRST quantization
\cite{AD1} hints at new and unexpected relations between the Hamiltonian
and Lagrangian BRST schemes.

\noi
In this paper we have focused on some of the more mathematical aspects
of the theory of higher brackets. The formulation has been greatly
simplified, thereby providing a much cleaner setting for the field theory
aspects. Of course, one most interesting result is the close correspondence
between higher antibrackets and the so-called string products of closed
string field theory \cite{Zwiebach}. We have argued that there are many
hints at the existence of a new product of string fields by means of which
non-polynomial closed string field theory could have at its origin a
formulation based on exponentials (defined within this product). This
remains speculation at the present stage, but even if it should turn out not
to be possible to realize such a product in closed string field theory,
our formalism may still be of use in this context. Namely, one may
conjecture that at least all those results which can be expressed solely
with the help of brackets (or, here, string products) may still be valid
in closed string field theory. Then the product may be used only in
intermediate steps, to simplify the calculations.

\noi
The BRST symmetry associated with higher antibrackets is part of a more
general BRST--anti-BRST symmetry, and we have shown how they both can be
included in a manifestly $Sp(2)$-covariant formulation. As an amusing
by-product of this, we can also write down  $Sp(2)$-covariant
analogues of the closed string field theory equations of motions, and
the corresponding $Sp(2)$-extended gauge symmetries. For the path integral
of conventional quantum field theory, the associated $Sp(2)$-covariant
BRST symmetry is required when one imposes certain identities as
$Sp(2)$-BRST Ward Identities in the path integral, as discusses in
the analogous case without $Sp(2)$ symmetry in ref. \cite{AD2}. It is
interesting that this $Sp(2)$-covariant formulation in a most natural
manner arises from the mathematical structure of strongly homotopy
Lie algebras.

\vspace{0.5cm}

\noindent
{\sc Acknowledgement:}~ We thank J. Stasheff and F. Akman for a
stimulating exchange of e-mails. We gratefully acknowledge discussions
with K. Suehiro and B. Zwiebach about the difficulties involved when
trying to identify directly the higher antibrackets with the string 
products of closed string field theory.
The work of KB and PHD is partially supported
by NorFA grants no.  95.30.182-O and 96.15.053-O, respectively. The work
of J.A. is partially supported by Fondecyt 1950809 and a
CNRS-CONICYT collaboration.

\vspace{0.5cm}

\begin{appendix}
\setcounter{equation}{0}
\section{Generalizations}

\noi
In section 2.2 we discussed the construction of higher antibrackets
based on the co-multiplication $\lambda$ of eq. (\ref{Koszullambda}).
Interestingly, this lends
itself to a natural generalization which we will outline in this appendix.
It is based on a
map  of degree $k$ called
$\lambda^{\epsilon_{1} ,\ldots, \epsilon_{k}}_{t_{1} ,\ldots, t_{k}} :
S{\cal A} \rightarrow \prod_{i=1}^k S{\cal A}$. Define it as follows:
\bea
\lefteqn{ \lambda^{\epsilon_{1} ,\ldots, \epsilon_{k}}_{t_{1} ,\ldots, t_{k}}
(1) = (1, 1, \ldots,1) } \cr && \cr
\lefteqn{ \lambda^{\epsilon_{1} ,\ldots, \epsilon_{k}}_{t_{1} ,\ldots, t_{k}}
(A_1 \otimes \ldots \otimes A_n)   }  \cr &=& \left(t_1(A_1 , 1, \ldots,1)
+(-1)^{\epsilon_{A_1}\epsilon_{2}}t_2 (1 , A_1 , 1, \ldots,1)+ \ldots
+(-1)^{\epsilon_{A_1} (\epsilon_{2}+ \ldots+\epsilon_{k})}t_k(1, \ldots,1,A_1)
\right) \cr && \cr &&\otimes \ldots \otimes \cr && \cr
&& \left(t_1(A_n, 1, \ldots,1)
+(-1)^{\epsilon_{A_n}\epsilon_{2}}t_2 (1 , A_n , 1, \ldots,1)+ \ldots
+(-1)^{\epsilon_{A_n} (\epsilon_{2}+ \ldots+\epsilon_{k})}t_k(1,
\ldots,1,A_n)\right) ~.
\eea

\noi
Here $\epsilon_{1} ,\ldots, \epsilon_{k}$ are $\pm 1$,
and $t_1, \ldots, t_k$ complex numbers.
Even though $\lambda^{\epsilon_{1} ,\ldots, \epsilon_{k}}_{t_{1} ,\ldots,
t_{k}} $ does not depend on $\epsilon_{1}$, it is natural to introduce an
$\epsilon_{1}$. Here $\prod_{i=1}^k S{\cal A}$ is equipped with a graded
product $\otimes$:
\beq
(A_1, \ldots,A_k) \otimes  (B_1, \ldots,B_k)
= (-1)^{\sum_{i>j}\epsilon_{A_i} \epsilon_{B_j} }
(A_1 \otimes B_1, \ldots,A_k\otimes B_k) ~.
\eeq

\noi
It turns out to be more convenient allowing for tensor valued operator $T$ ,
i.e.\ linear maps: $S{\cal A} \rightarrow S{\cal A}$
instead of just working with ordenary linear
operators: ${\cal A} \rightarrow {\cal A}$.
We can now define generalized higher brackets
for  \mbox{$T_1, \ldots, T_k  \in {\rm Hom}_{C}(S{\cal A},S{\cal A})$}
and complex numbers $t_1, \ldots, t_k$
\beq
\left| \matrix{
T_1 & \cdots & T_k\cr
t_1 & \cdots & t_k \cr
}\right|
\equiv  (T_1 \times \ldots \times T_k) \circ
\lambda^{\epsilon_{T_1} ,\ldots, \epsilon_{T_k}}_{t_{1} ,\ldots, t_{k}}
 \in {\rm Hom}_{C}(S{\cal A},S{\cal A}) ~.
\eeq
respectively
\beq
\left\{ \matrix{
T_1 & \cdots & T_k\cr
t_1 & \cdots & t_k \cr
}\right|
\equiv \left| \matrix{
T_1 & \cdots & T_k\cr
t_1 & \cdots & t_k \cr
}\right|^{\sim}
 \in {\rm Hom}_{C}(S{\cal A},{\cal A}) ~.
\eeq

\noi
With the above generalization, we can write
(cf.\  (\ref{Koszullambda}), (\ref{PhiT}), (\ref{PhinT}) resp.
(\ref{invPhiT}).\
 ) :
\bea
\lambda &=& \lambda^{\epsilon_T,~1}_{1, -1} \cr
&& \cr
\Phi_T &=& \left| \matrix{
T &  {\rm Id}_{S{\cal A}} \cr
1  & -1 \cr
}\right| \cr
&& \cr
\widetilde{\Phi}_T &=& \left\{\matrix{
T &  {\rm Id}_{S{\cal A}} \cr
1  & -1 \cr
}\right| \cr
&& \cr
 T &=& \left| \matrix{
\Phi_T &  {\rm Id}_{S{\cal A}} \cr
1  & 1 \cr
}\right| \ = \
 \left\{\matrix{
\widetilde{\Phi}_T &  {\rm Id}_{S{\cal A}} \cr
1  & 1 \cr
}\right| ~.
\eea

\noi
The first few brackets can be rewritten as
\bea
 \left. \left|   \matrix{
 T \cr
t   \cr
} \right|
A_1 \otimes \ldots \otimes A_n \right\}
&=&  t^n \ T( A_1 \otimes \ldots \otimes A_n)
\label{Tt} \\
 \left. \left|   \matrix{
S & T  \cr
s & t   \cr
} \right|
A_1 \otimes \ldots \otimes A_n \right\}
&=&  (S \otimes T)
\left( (\widehat{sA_1}^T+\widehat{tA_1}^S )
\otimes \ldots \otimes
(\widehat{sA_n}^T+\widehat{tA_n}^S )
\right)  \cr
&=&\sum_{\ell_1, \ldots ,\ell_n =0}^1
 (-1)^{\epsilon_T \sum_j \epsilon_{A_j}  \ell_j
+\sum_{j>k}\epsilon_{A_j} \epsilon_{A_k} \ell_j (1-\ell_k) }
 S((sA_1)^{\ell_1} \otimes \ldots \otimes (sA_n)^{\ell_n})  \cr
&& ~~~~~~~~~~~~~~\otimes \
T ((tA_1)^{1-\ell_1} \otimes \ldots \otimes (tA_n)^{1-\ell_n} )
\label{TtSs}
\eea
\bea
\lefteqn{   \left. \left|   \matrix{
S & T &  U \cr
s & t  & u \cr
} \right|
A_1 \otimes \ldots \otimes A_n \right\}  } \cr
&=&  (S \otimes T \otimes U)
\left( (\widehat{sA_1}^{TU}+\widehat{tA_1}^{SU}+\widehat{uA_1}^{S T}  )
\otimes \ldots \otimes
 (\widehat{sA_n}^{TU}+\widehat{tA_n}^{SU}+\widehat{uA_n}^{S T}  )
\right)  \cr
&=& \sum_{\ell_1, \ldots ,\ell_n =-1}^1
(-1)^{(\epsilon_T+\epsilon_U) \sum_j \epsilon_{A_j}  \ell_j^{-}
+ \epsilon_U \sum_j \epsilon_{A_j} (1-|\ell_j |)
+\sum_{j>k} \epsilon_{A_j} \epsilon_{A_k}  \left( \ell_j^{-}
(1-| \ell_k|)
+ (1-| \ell_j|)  \ell_k^{+} + \ell_j^{-}  \ell_k^{+}   \right)  } \cr
&& ~~~~~~~~~~~~~~
 S((sA_1)^{\ell_1^{-}} \otimes \ldots \otimes  (sA_n)^{\ell_n^{-}})
\otimes  \ T((tA_1)^{1-| \ell_1| }  
\otimes \ldots \otimes  (tA_n)^{1-| \ell_n|
}) \cr
&& ~~~ ~~~~~~~~~~~~~~~~~~~~~~~~~\otimes \
 U((uA_1)^{\ell_1^{+}}  \otimes \ldots \otimes  (uA_n)^{\ell_n^{+}}) ~.
\label{TtSsUu}
\eea
Here $\ell^{\pm}={1 \over 2}( |\ell | \pm \ell)$
is just the positive (resp.\ negative) part of the real number $\ell$.
This is clearly not a systematical way of describing the generalized
brackets for more than three operator entries.
In order to proceed into higher numbers of operator entries,
we use characteristic functions.
Let $\chi_{{}_{\scriptstyle s}}$ be the characteristic function 
associated with the statement $s$.
$\chi_{{}_{\scriptstyle {s}}}=1$ if $s$ is true, and $\chi_{{}_{\scriptstyle
s}}=0$ if $s$ is false.
Then
\bea
 \lefteqn{   \left.\left| \matrix{
T_1 & \cdots & T_k\cr
t_1 & \cdots & t_k \cr
}\right|
A_1 \otimes \ldots \otimes A_n \right\}  } \cr
&=&  (T_1 \otimes \ldots \otimes T_k)
\left( (\widehat{t_1 A_1}^{T_2\ldots T_k}
+\widehat{t_1 A_1}^{T_1T_3\ldots T_k}
+\ldots +\widehat{t_1 A_1}^{T_1 \ldots T_{k-1}}  ) \right. \cr
&& ~~~~~~~~~~~~~~ \left. \otimes \ldots \otimes
(\widehat{t_n A_n}^{T_2\ldots T_k}
+\widehat{t_n A_n}^{T_1T_3\ldots T_k}
+\ldots+\widehat{t_n A_n}^{T_1 \ldots T_{k-1}}  )
\right)  \cr
&=& \sum_{\ell_1, \ldots ,\ell_n =1}^k
(-1)^{ \sum_{r>s} \sum_p \epsilon_{T_r} \epsilon_{A_p}\chi_{{}_{\scriptstyle
\ell_p=s}}
+ \sum_{p>q} \epsilon_{A_p} \epsilon_{A_q}
 \sum_{r<s} \chi_{{}_{\scriptstyle \ell_p=r}} 
\chi_{{}_{\scriptstyle \ell_q=s}}
} \cr
&& ~~~~~~~~~~~~~~
 T_1((t_1 A_1)^{\chi_{{}_{\scriptstyle \ell_1=1}}}
\otimes \ldots \otimes  (t_n A_n)^{\chi_{{}_{\scriptstyle \ell_n=1}}} ) \cr
&& ~~~~~~~~~~~~~~ \otimes \ldots \otimes
 T_k((t_1 A_1)^{\chi_{{}_{\scriptstyle \ell_1=k}}}
\otimes \ldots \otimes  (t_n A_n)^{\chi_{{}_{\scriptstyle \ell_n=k}}} ) ~,
\eea

\noi
The generalized higher antibrackets are all graded symmetric:
\beq
\left| \matrix{
T_1 & \cdots & T_k\cr
t_1 & \cdots & t_k \cr
}\right|
=  (-1)^{\epsilon_{\tau}  }
\left | \matrix{
T_{\tau(1)} & \cdots & T_{\tau(k)}\cr
t_{\tau(1)} & \cdots & t_{\tau(k)} \cr
}\right| ~,
\eeq
where $(-1)^{\epsilon_{\tau}}$ is the sign factor
originating from permuting Grassmann graded quantities:
\beq
   (T_1,\ldots,T_k)
\mapsto (T_{\tau(1)} ,\ldots,T_{\tau(k)} ) ~.
\eeq
They are restricted linear and enjoy simple composition properties:
\bea
\left| \matrix{
T_1 & \cdots &T_i'+T_i'' & \cdots & T_k\cr
t_1 & \cdots &t_i& \cdots & t_k \cr
}\right|
&=& \left| \matrix{
T_1 & \cdots &T_i' & \cdots & T_k\cr
t_1 & \cdots &t_i& \cdots & t_k \cr
}\right|
+\left| \matrix{
T_1 & \cdots &T_i'' & \cdots & T_k \cr
t_1 & \cdots &t_i& \cdots & t_k \cr
}\right| \cr && \cr
\left| \matrix{
T_1 & \cdots &T_i & \cdots & T_k\cr
t_1 & \cdots &t_i'+t_i''& \cdots & t_k \cr
}\right|
&=& \left| \matrix{
T_1 & \cdots &T_i &T_i &  \cdots & T_k\cr
t_1 & \cdots &t_i'&t_i''& \cdots & t_k \cr
}\right|  \cr && \cr
\left| \matrix{
T_1 & \cdots & T_k\cr
t t_1 & \cdots  & t t_k \cr
}\right|
&=&t^k \left| \matrix{
T_1  &  \cdots & T_k\cr
t_1 & \cdots & t_k \cr
}\right|  \cr && \cr
\left| \matrix{
 \left| \matrix{
S_1  &  \cdots & S_k\cr
s_1 & \cdots & s_k \cr
}\right|
&T_1 & \cdots & T_{\ell}\cr
 s&t_1 & \cdots  &  t_{\ell} \cr
}\right|
&=& \left| \matrix{
S_1  &  \cdots & S_k&T_1  &  \cdots & T_{\ell}\cr
s s_1 & \cdots & s s_k&t_1 & \cdots & t_{\ell} \cr
}\right| ~.
\eea
When all the coefficients $t_1, \ldots, t_k=1$ are equal to $1$,
one can say a lot more.
First of all let us simplify the notation in this special case:
\beq
\left| T_1 , \ldots , T_k  \right| \equiv
\left| \matrix{
T_1 & \cdots & T_k\cr
t_1=1 & \cdots & t_k=1 \cr
}\right| ~.
\label{specdef}
\eeq
Following Zwiebach (\cite{Zwiebach}, eq.\  (4.100)),
we defines a co-derivation $b_T$ for an operator
\mbox{$T \in {\rm Hom}_{C}(S{\cal A},S{\cal A})$.}
\beq
 b_T \equiv \left| T ,  {\rm Id}_{S{\cal A}} \right| \equiv
\left| \matrix{
T &  {\rm Id}_{S{\cal A}} \cr
1  & 1 \cr
}\right|
\label{bTdef1}
\eeq
\bea
  \lefteqn{ b_T(A_1 \otimes \ldots \otimes A_n) = T((A_1+
\widehat{A_1}^T) \otimes \ldots \otimes (A_n +\widehat{A_n}^T))} \cr
&=& \sum_{i_1, \ldots ,i_n =0}^1 (-1)^{\sum_{j>k}\epsilon_{A_j}
\epsilon_{A_k} i_j (1-i_k) }T(A_1^{i_1} \otimes \ldots \otimes A_n^{i_n})
 \otimes A_1^{1-i_1} \otimes \ldots \otimes A_n^{1-i_n} ~.
\label{bTdef2}
\eea
We also note the following simple relations
\bea
b_{S+T}&=&b_S + b_T \cr
\pi_{\cal A} \circ b_T &=& T \cr
b_{\textstyle \Phi_T }&=& T \ = \
\widetilde{b}_{\textstyle \widetilde{\Phi}_T}
\label{bsim}
\eea
The two last statements only holds for operator $T$,
which  is not tensor valued, i.e.\
\mbox{$T \in {\rm Hom}_{C}(S{\cal A},{\cal A})$.}
Less obvious are the following identities:
\bea
b_S \circ b_T &=& b_{S{\textstyle  \circ b_T}} + b_{\left| S, T \right| } \cr
&&
\cr
\left| S_1, \ldots, S_k \right| \circ b_T
&=&\sum_{i=1}^k
(-1)^{\epsilon_{T} \left( \epsilon_{S_{i+1}}+ \ldots +\epsilon_{S_k} \right) }
\left| S_1, \ldots, S_i \circ b_T, \ldots, S_k \right|~,
\label{bder}
\eea
where \mbox{$T \in {\rm Hom}_{C}(S{\cal A},{\cal A})$.}
The first identity in (\ref{bder}) is an important special
case of the second identity.
Many of the above (and coming) constructions
can actually be carried out in a vector space frame just as well,
i.e.\ not assuming a dot product for the algebra ${\cal A}$.
For instance the $\Phi_T$ and $b_T$
construction works without a dot,
if \mbox{$T \in {\rm Hom}_{C}(S{\cal A},S{\cal A})$.}
The most notable exceptions are the tilde operation,
the higher brackets $\Phi^n_T$, and in particular the recursion relation
(\ref{recur}).
However, one can impose the existence of the higher brackets $\Phi^n_T$
(and their so-called ``main identity'': see below)
as a principle. For instance in closed string field theory
the higher brackets can be built up
from a geometric consideration on moduli space
\cite{Zwiebach}.

\setcounter{equation}{0}
\section{\sc Higher Main identities}

\noi
The purpose of this appendix is to show that
by applying the lemma (\ref{lemma}) and (\ref{bder}) several times,
one can derive higher order versions of the same lemma.
Unfortunately, there is no closed expression
for \mbox{$\Phi_{T_1 T_2 \ldots T_k}$} in terms of higher brackets
\mbox{$\Phi_{T_1},\widetilde{\Phi}_{T_2} ,
\ldots ,\widetilde{\Phi}_{T_k}$} alone,
but there is a fairly simple graphical representation, which we now
sketch.

\noi
We will argue that \mbox{$\Phi_{T_1 T_2 \ldots T_k}$} can be understood as
a restricted sum over oriented and connected tree diagrams 
with $k$~ 1-, 2- and 3-vertices.

\noi
First of all, we take every line in the tree to run between vertices.
In particular: every external leg is assumed decorated with an
external point, a ``1-vertex''.
All other vertices but a {\em root}-vertex are supposed to have at least one
in-going line.
Because there are at most three lines connected to each vertex,
one can draw all oriented lines in the tree
horizontally downwards, and vertically to the right.
\begin{itemize}
\item
Each vertex corresponds to a higher bracket $\widetilde{\Phi}_{T_i}$.
\item
A horizontally connected collection of $r-1$ oriented lines
$r=1,2,3,\ldots$,
correspons to a
$r$-{\em bracket-bracket}
\mbox{$\left| \widetilde{\Phi}_{T_{i_1}},
\ldots ,\widetilde{\Phi}_{T_{i_r}} \right| $},
where \mbox{$i_1 <i_2 < \ldots <i_r$} (cf. definition (\ref{specdef})).
Of course one can skip the horisontal orientation
\mbox{$i_1 <i_2 < \ldots <i_r$} inside a  bracket-bracket,
at the cost of introducing a symmetry factor
$\frac{1}{r!}$ for each bracket-bracket.
\item
A downward line corresponds to the action
of the co-derivation $(\cdot) \circ b_{(\cdot\cdot)}$
\beq
\widetilde{\Phi}_{T_i} \circ b_{\textstyle \left| \widetilde{\Phi}_{T_{i_1}} ,
\ldots ,\widetilde{\Phi}_{T_{i_r}} \right| } ~,
\eeq
with $i<i_1, \ldots,i<i_r$.
(A conventional higher bracket  \mbox{$\widetilde{\Phi}_{T_i}=\left|
\widetilde{\Phi}_{T_i} \right| $}
is also considered to be a 1-bracket-bracket.)
An incoming  downward lines actual attachment position to a bracket-bracket 
is immaterial, and tree diagram with different incoming attachment position
are considered equal, and should only be counted as one.
\item
Each tree is given a sign, because of the permutation of Grassmann-graded
brackets within it.
The easiest way to specify this sign is to enumerate the vertices, which is
basically the same as specifying a permutation $\tau \in {\cal S}_k$ that
takes the enumeration of the operators
\mbox{$T_1, T_2 ,\ldots ,T_k$} into this enumeration of the vertices.
The sign is then computed as the sign
originating from simply permuting Grassmann graded quantities:
\beq
   (T_1,\ldots,T_k)
\mapsto (T_{\tau(1)} ,\ldots,T_{\tau(k)} ) ~.
\eeq
The vertex enumeration goes as follows:
Start at the left-uppermost vertex, proceed downwards if possible, else
to the right. When entering a bracket-bracket, start with the left entry.
When hitting an end-bracket-bracket, go back to the last furcation point
(that is, the next-to-last bracket-bracket), then go to the right, etc.
\end{itemize}

\noi
\underline{Proof:} (sketched here only for the bosonic case).
We use induction in the total number $k$
of $1$-,$2$- and $3$-vertices. From the lemma
\beq
 \Phi_{T_1 T_2 \ldots T_{k+1}} =
\Phi_{T_1 T_2 \ldots T_k} \circ b_{\widetilde{\Phi}_{T_{k+1}} }
+ \left| \Phi_{T_1 T_2 \ldots T_k}, \widetilde{\Phi}_{T_{k+1}} \right|
\label{mainidinduc}
\eeq
Now each tree with $k+1$ vertices of the above type
can be grown from a tree with $k$ vertices
by attaching either an extra \mbox{$\widetilde{\Phi}_{T_{k+1}}$}-entry
to the right (which gives a horizontal growth) in a bracket-bracket,
or a downward growth \mbox{$\circ b_{\widetilde{\Phi}_{T_{k+1}} } $},
from a vertex, if there is not already an outgoing downward line there.
It is easy to see from (\ref{bder}) that the action of
\mbox{$\circ b_{\widetilde{\Phi}_{T_{k+1}} } $} on all the trees with
$k$ vertices \mbox{$\Phi_{T_1 T_2 \ldots T_k}$}
yields all the trees with $k+1$ vertices exactly once,
except the diagram where the {\em root}-bracket-bracket
is enlarged by an entry to the right.
This tree is then built via the second term
on the right hand side of (\ref{mainidinduc}).

\begin{flushright}
${\,\lower0.9pt\vbox{\hrule \hbox{\vrule
height 0.2 cm \hskip 0.2 cm \vrule height 0.2 cm}\hrule}\,}$
\end{flushright}

\noi
To make this construction more tangible, let us evaluate some of the
lowest cases:
\beq
 \Phi_{ST} = \Phi_S \circ b_{\textstyle \widetilde{\Phi}_{T} }
+ \left| \Phi_{S}, \widetilde{\Phi}_{T} \right|
\eeq
\bea
\Phi_{STU} &=& \Phi_{S} \circ b_{ \textstyle   \widetilde{\Phi}_{T}
\circ b_{ \textstyle   \widetilde{\Phi}_{U} } }
+  \Phi_{S} \circ b_{ \textstyle  \left|   \widetilde{\Phi}_{T} ,
  \widetilde{\Phi}_{U} \right| }
+  \left|   \Phi_{S} \circ b_{ \textstyle  \widetilde{\Phi}_{T}} ,
 \widetilde{\Phi}_{U} \right| \cr
&&\cr &&+ (-1)^{\epsilon_{T}\epsilon_{U}} \left|
\Phi_{S} \circ b_{ \textstyle  \widetilde{\Phi}_{U}} ,
  \widetilde{\Phi}_{T} \right|
+  \left|   \Phi_{S}  ,
  \widetilde{\Phi}_{T} \circ b_{ \textstyle  \widetilde{\Phi}_{U}} \right|
+ \left| \Phi_{S} , \widetilde{\Phi}_{T} ,  \widetilde{\Phi}_{U} \right| ~.
\label{hmainid3}
\eea
\bea
\Phi_{STUV} &=& \Phi_{S} \circ b_{ \textstyle   \widetilde{\Phi}_{T}
\circ b_{ \textstyle   \widetilde{\Phi}_{U}\circ
b_{ \textstyle   \widetilde{\Phi}_{V} } } }
+\Phi_{S} \circ b_{ \textstyle   \widetilde{\Phi}_{T}
 \circ b_{ \textstyle  \left|   \widetilde{\Phi}_{U} ,
  \widetilde{\Phi}_{V} \right| }}
+ \left|  \Phi_{S} \circ b_{ \textstyle   \widetilde{\Phi}_{T}
 \circ b_{ \textstyle   \widetilde{\Phi}_{U} } },
  \widetilde{\Phi}_{V} \right|  \cr
&&\cr &&+ (-1)^{\epsilon_{U}\epsilon_{V}}  \left|
\Phi_{S} \circ b_{ \textstyle   \widetilde{\Phi}_{T}
 \circ b_{ \textstyle   \widetilde{\Phi}_{V} } },
  \widetilde{\Phi}_{U} \right|
+ (-1)^{\epsilon_{T}(\epsilon_{U}+\epsilon_{V})}
\left|  \Phi_{S} \circ b_{ \textstyle   \widetilde{\Phi}_{U}
 \circ b_{ \textstyle   \widetilde{\Phi}_{V} } },
  \widetilde{\Phi}_{T} \right| \cr
&&\cr &&+\left|  \Phi_{S} ,
  \widetilde{\Phi}_{T} \circ b_{ \textstyle   \widetilde{\Phi}_{U}
 \circ b_{ \textstyle   \widetilde{\Phi}_{V} } } \right|
+\Phi_{S} \circ b_{ \textstyle   \left|  \widetilde{\Phi}_{T}
 ,  \widetilde{\Phi}_{U} ,
  \widetilde{\Phi}_{V} \right| }
+\left|  \Phi_{S} \circ b_{ \textstyle   \left|  \widetilde{\Phi}_{T}
 ,  \widetilde{\Phi}_{U} \right| } ,
  \widetilde{\Phi}_{V} \right| \cr
&&\cr &&+ (-1)^{\epsilon_{U}\epsilon_{V}}
\left|  \Phi_{S} \circ b_{ \textstyle   \left|  \widetilde{\Phi}_{T}
 ,  \widetilde{\Phi}_{V} \right| } ,
  \widetilde{\Phi}_{U} \right|
+ (-1)^{\epsilon_{T}(\epsilon_{U}+\epsilon_{V})}
\left|  \Phi_{S} \circ b_{ \textstyle   \left|  \widetilde{\Phi}_{U}
 ,  \widetilde{\Phi}_{V} \right| } ,
  \widetilde{\Phi}_{T} \right| \cr
&&\cr &&+\left|  \Phi_{S},
  \widetilde{\Phi}_{T} \circ b_{ \textstyle   \left|  \widetilde{\Phi}_{U}
 ,  \widetilde{\Phi}_{V} \right| }  \right|
+\left|  \Phi_{S} \circ b_{ \textstyle   \widetilde{\Phi}_{T} }
 ,  \widetilde{\Phi}_{U} ,
  \widetilde{\Phi}_{V} \right|
+(-1)^{\epsilon_{T}\epsilon_{U}}
\left|  \Phi_{S} \circ b_{ \textstyle   \widetilde{\Phi}_{U} }
 ,  \widetilde{\Phi}_{T} ,
  \widetilde{\Phi}_{V} \right| \cr
&&\cr &&+(-1)^{(\epsilon_{T}+\epsilon_{U})\epsilon_{V}}
\left|  \Phi_{S} \circ b_{ \textstyle   \widetilde{\Phi}_{V} }
 ,  \widetilde{\Phi}_{T} ,
  \widetilde{\Phi}_{U} \right|
+\left|  \Phi_{S} ,
\widetilde{\Phi}_{T} \circ b_{ \textstyle   \widetilde{\Phi}_{U} } ,
  \widetilde{\Phi}_{V} \right| \cr
&&\cr &&+ (-1)^{\epsilon_{U}\epsilon_{V}} \left|  \Phi_{S} ,
\widetilde{\Phi}_{T} \circ b_{ \textstyle   \widetilde{\Phi}_{V} } ,
  \widetilde{\Phi}_{U} \right|
+\left|  \Phi_{S} ,  \widetilde{\Phi}_{T} ,
  \widetilde{\Phi}_{U} \circ b_{ \textstyle   \widetilde{\Phi}_{V} }  \right|
+\left|  \Phi_{S} ,  \widetilde{\Phi}_{T} ,
  \widetilde{\Phi}_{U},  \widetilde{\Phi}_{V}   \right|
{}~.
\eea

\noi
It is clear that these generalized (higher) main identities quickly
become totally unwieldy when written out in full. In the
special case of just one Grassmann-odd operator $T$,
the higher main identities actually give no genuinely new information
when $T^2 = 0$.
This nilpotent case can be seen using the graphical representation,
where  the main identity (\ref{lemma})
states that one vertical line (with a bracket at each end) is equal to zero.
This means that an end-bracket-bracket
that contains precisely one bracket causes the tree to vanish.
Any end-bracket-bracket containing more brackets causes the tree to vanish,
because the brackets are Grassmann odd. However, already in the simple
case of just one odd operator $T$ which is {\em not} nilpotent, the above
generalized main identities relate brackets based on $T^n$ (up to as many
powers possible while still having $T^n \neq 0$) to those based on $T$.
Koszul \cite{Koszul} has given one particular example of these identities,
but no general prescription for finding them.

\noi
The case of two anticommuting odd
operators $T_{1}$ and $T_{2}$ is even more interesting.
Let us restate the main identity (\ref{TSmainid}) as
\beq
0 =  \widetilde{\Phi}_{T_{ \{ a}} \circ b_{ \textstyle
\widetilde{\Phi}_{T_{b   \}  }  } }
\eeq
At the next level, one new main identity arises:
\bea
 0 &=&  \widetilde{\Phi}_{T_a} 
\circ b_{ \textstyle   \widetilde{\Phi}_{T_b}
\circ b_{ \textstyle   \widetilde{\Phi}_{T_c} } }
+ \widetilde{\Phi}_{T_a} \circ b_{ \textstyle  
\left|   \widetilde{\Phi}_{T_b} ,
  \widetilde{\Phi}_{T_c} \right| }
\label{newmain}
\eea
Shown in more details, the content of this new main identity
involving two operators $T_{1}$ and $T_{2}$ is:
\bea
0 &=& \sum_{r,s=0}^{r+s \leq n} {1\over {r! s! (n-r-s)! }}
\sum_{\pi \in {\cal S}_n } (-1)^{\epsilon_{\pi}}
\Phi^{n-r-s+1}_{T_a}\left(  \Phi_{T_b}^{s+1} \left(
\Phi^r_{T_c} (A_{\pi(1)} \otimes \ldots \otimes A_{\pi(r)}  ) \right. \right.
\cr
&&~~~~~~~~~~~~~~~~~~~~~ \left.  \left.
\otimes \ A_{\pi(r+1)} \otimes \ldots \otimes A_{\pi(r+s)} \right)
 \otimes A_{\pi(r+s+1)} \otimes \ldots \otimes A_{\pi(n)} \right)  \cr &&\cr
&& +\sum_{r,s=0}^{r+s \leq n} {1\over {r! s! (n-r-s)! }}
\sum_{\pi \in {\cal S}_n }
(-1)^{\epsilon_{\pi}+\epsilon_{A_{\pi(1)}}
+ \ldots +\epsilon_{A_{\pi(r)}} }
\Phi^{n-r-s+2}_{T_a}\left(  \Phi_{T_b}^{r} (A_{\pi(1)} \otimes \ldots \otimes
A_{\pi(r)}
) \right. \cr
&&~~~~~~~~~~~~~~~~~~~~~ \left. \otimes \
 \Phi^s_{T_c} (A_{\pi(r+1)} \otimes \ldots \otimes A_{\pi(r+s)} ) \otimes
A_{\pi(r+s+1)} \otimes \ldots \otimes A_{\pi(n)} \right)
\eea
Remarkably, this identity holds without any kind of
symmetrisation in the indices \mbox{$a,b,c=1,2$}.
Only the case \mbox{$b \neq c$} is truly a new identity.
{}For other combinations of \mbox{$a,b,c=1,2$}, eq.\ (\ref{newmain})
can be deduced from the original main identity and
symmetry arguments. One can prove that all nilpotent $Sp(2)$-symmetric
higher main identities can be derived from these
two main identities.  This follows from:
\bea
0 &=&  \left|   \widetilde{\Phi}_{T_a} , \widetilde{\Phi}_{T_b} , 
 \widetilde{\Phi}_{T_c} , \ldots \right|
\eea
\bea
 0 &=& \sum_{{\rm cycl.\ }a,b,c~}
 \left|   \widetilde{\Phi}_{T_a} 
\circ b_{ \textstyle  \widetilde{\Phi}_{T_b}} ,
 \widetilde{\Phi}_{T_c}, \ldots \right| ~.
\eea
The analogous 
$Sp(2)$-symmetric formulations have been discussed in section 4.

\end{appendix}

\end{document}